\documentclass[11pt,twocolumn,tight,times]{aastex631}
\usepackage{graphicx,amsmath}
\usepackage{color}
\usepackage{booktabs}
\usepackage{bm}

\usepackage{times}
\usepackage{float}
\usepackage{ulem}
\usepackage{multirow}

\newcommand{\adsurl}[1]{\href{#1}{ADS}}
\providecommand{\url}[1]{\href{#1}{#1}}

\newcommand{\be}{\begin{equation}}
\newcommand{\ee}{\end{equation}}
\newcommand{\bea}{\begin{eqnarray}}
\newcommand{\eea}{\end{eqnarray}}

\usepackage{color}
\usepackage{ifthen}
\newboolean{editorial}
\setboolean{editorial}{true}
\newcommand{\editorial}[2]{\ifthenelse{\boolean{editorial}}{\textcolor{red}{[\textsf{\textbf{{#1}}}: }\textcolor{blue}{\textsf{{#2}}}\textcolor{red}{]}}{}}

\shorttitle{Localization, Host-identification, and Cosmological-Inference of Lensed GW events}
\shortauthors{Chen et al.}

\begin{document}
\title{Enhanced Localization of Dark Lensed Gravitational Wave Events Enables Host Galaxy Identification and Precise Cosmological Inference }
\author[0000-0001-7952-7945]{Zhiwei Chen}
\affiliation{National Astronomical Observatories, Chinese Academy of Sciences, 20A Datun Road, Beijing 100101, China}
\affiliation{School of Astronomy and Space Sciences, University of Chinese Academy of Sciences, 19A Yuquan Road, Beijing 100049, China}
\author[0000-0002-1745-8064]{Qingjuan Yu}\thanks{yuqj@pku.edu.cn}
\affiliation{Kavli Institute for Astronomy and Astrophysics, and School of Physics, Peking University, Beijing 100871, People's Republic of China}
\author[0000-0002-1310-4664]{Youjun Lu}\thanks{luyj@nao.cas.cn}
\affiliation{National Astronomical Observatories, Chinese Academy of Sciences, 20A Datun Road, Beijing 100101, China}
\affiliation{School of Astronomy and Space Sciences, University of Chinese Academy of Sciences, 19A Yuquan Road, Beijing 100049, China}
\author[0000-0001-5174-0760]{Xiao Guo}
\affiliation{Institute for Gravitational Wave Astronomy, Henan Academy of Sciences, Zhengzhou 450046, Henan, China}

\begin{abstract}
Lensed gravitational wave (GW) events are expected to be powerful new probes of cosmology, contingent on redshift measurement by electromagnetic observations. Host galaxy identification is thus crucial but challenging due to poor localization by GW signal alone. In this paper, we show that the third-generation ground-based GW detectors will detect a population of lensed events with three or more detectable images (including the central one), each arriving at distinct times and Earth locations in the space, forming an effective network that reduces the typical localization area to $\sim0.01$\,deg$^2$. For at least $90\%$ (or $50\%$) of these events, the localization improves by more than a factor of $10$ (or $30$) comparing with unlensed cases. Such precise localization and multiple-image detections enable robust host-galaxy identification and, through lens modelling, further yield sub-arcsecond position. As ``dark lensed sirens", these events become powerful probes of cosmological parameters. Using simulated lensed compact-binary mergers, we show that two-year or longer observations with third-generation GW detectors can measure the Hubble constant to $\lesssim1$\% precision via ``dark lensed sirens" (even when relying solely on lensed stellar-mass binary black hole events), while simultaneously constraining other cosmological parameters. This approach will provide an independent, complementary avenue for measuring cosmological parameters.
\end{abstract}

\section{Introduction}
\label{sec:intro}

Since the first detection of gravitational wave (GW) signal emitted by stellar binary black hole merger, GW150914, more than two hundred of GW events have been detected by Laser Interferometer Gravitational wave Observatory (LIGO)-Virgo-KAGRA (LVK) network \citep[e.g., ][see also at \url{https://gracedb.ligo.org/}]{2016PhRvL.116m1102A,2019PhRvX...9c1040A,2020arXiv201014527A,2021arXiv211103606T,2021arXiv211103634T}. With future 2.5th and 3rd generation ground-based detectors, such as LIGO Voyager \citep{2020CQGra..37p5003A}, Einstein Telescope \citep[ET;][]{ETstudy} and Cosmic Explorer \citep[CE;][]{2019BAAS...51g..35R}, a large number of GW events can be detected. Among which a fraction will be strongly gravitationally lensed by intervening galaxies as electromagnetic wave signals \citep[e.g.,][]{2015JCAP...12..006D,2018MNRAS.476.2220L,2018PhRvD..97b3012N, 2018MNRAS.480.3842O,2021ApJ...921..154W,2023ApJ...953...36C,2023MNRAS.518.6183M}. These events may serve as independent probes to constrain the cosmological parameters, while strongly dependent on the identification of their associated lensed host galaxies \citep{2018LRR....21....3A,2020MNRAS.498.3395H,2020MNRAS.497..204Y}, which is however very challenging due to the poor localization by GW detectors. 

In this paper, we show that the ground-based GW detectors, detecting three or more images from a lensed event at different times and Earth locations in the space, form an effective detector network, leading to a localization error of $\lesssim0.01$\,deg$^2$ for a typical lensed event. This precise localization almost grants the identification of the lensed host galaxy and further sub-arcsecond localization of the event by reconstructing the lensed host. Based on this effective network, we demonstrate the ``dark lensed siren" method to constrain the cosmological parameters, which offers an independent and complementary way to solve the Hubble tension and constrain the nature of dark energy. Here, ``dark" means that no bright EM counterpart of the lensed GW event is detected. Nevertheless, even without EM emission, the GW event can still be associated with its lensed host-galaxy found by galaxy surveys via its precise localization, and its position can be further narrowed to sub-arcsecond area via lensing reconstruction. The ``dark lensed siren" here is different from the traditional “dark standard siren”, for which the ``dark" refers to GW events without identified host galaxies due to their poor localization and no bright EM counterparts, forcing to infer cosmological parameters from the statistical redshift distribution of all galaxies inside the localization error regions \citep[e.g.,][]{1986Natur.323..310S,2021ApJ...909..218A,2022RAA....22a5020C,2024MNRAS.535..961B}. Furthermore, we assess the cosmological inference capability of ``dark lensed siren" by generating mock GW events and Markov Chain Monte Carlo (MCMC) simulations and find that  the Hubble constant can be efficiently determined with a precision of $\lesssim1$\% with two year's detection of the third generation GW detectors, and the other cosmological parameters can be also well constrained. 

This paper is organized as follows. In Section~\ref{sec:network}, we introduce the basic principle of the effective network composed of multiple lensed images of GW signals and its enhancement on the localization precision. In Section~\ref{sec:dark}, we demonstrate the ``dark lensed siren" method on constraining the cosmological parameters and assess its capability. The conclusions are provided in Section~\ref{sec:con}.

\section{Enhanced Localization by Effective Network}
\label{sec:network}

For a lensed gravitational wave (GW) event, the arrival times of its different images can be different due to their different paths. The differences between the arrival times of these images (or time-delays) are typically on the order of days to months \citep[e.g.,][]{2018MNRAS.476.2220L}, which can be measured with a precision of $\lesssim1$\,ms by the ground-based GW detectors. The lensed GW events detected by future GW observations may be applied to constrain the Hubble constant to a high precision by using the measurements of their time-delay ($\tau$) distances $D_{\tau}\equiv(1+z_{\ell})D_{\ell}D_{\rm s}/D_{\ell\rm s}$ and luminosity distances $d_{\rm L}$, if the redshifts of their electromagnetic (EM) counterparts/host galaxies can be independently obtained by multiwavelength EM observations \citep[e.g.,][]{2017NatCo...8.1148L,2020MNRAS.498.3395H}. Here $D_\ell$, $D_{\rm s}$, and $D_{\ell\rm s}$ denote the distances to the lens, source, and that between lens and source, respectively, and $z_{\ell}$ is the redshift of the lens. However, the poor localization of normal GW events solely given by GW signals may hinder the search for their EM counterparts/host galaxies.

\begin{figure}[ht]
\centering
\includegraphics[width=0.43\textwidth]{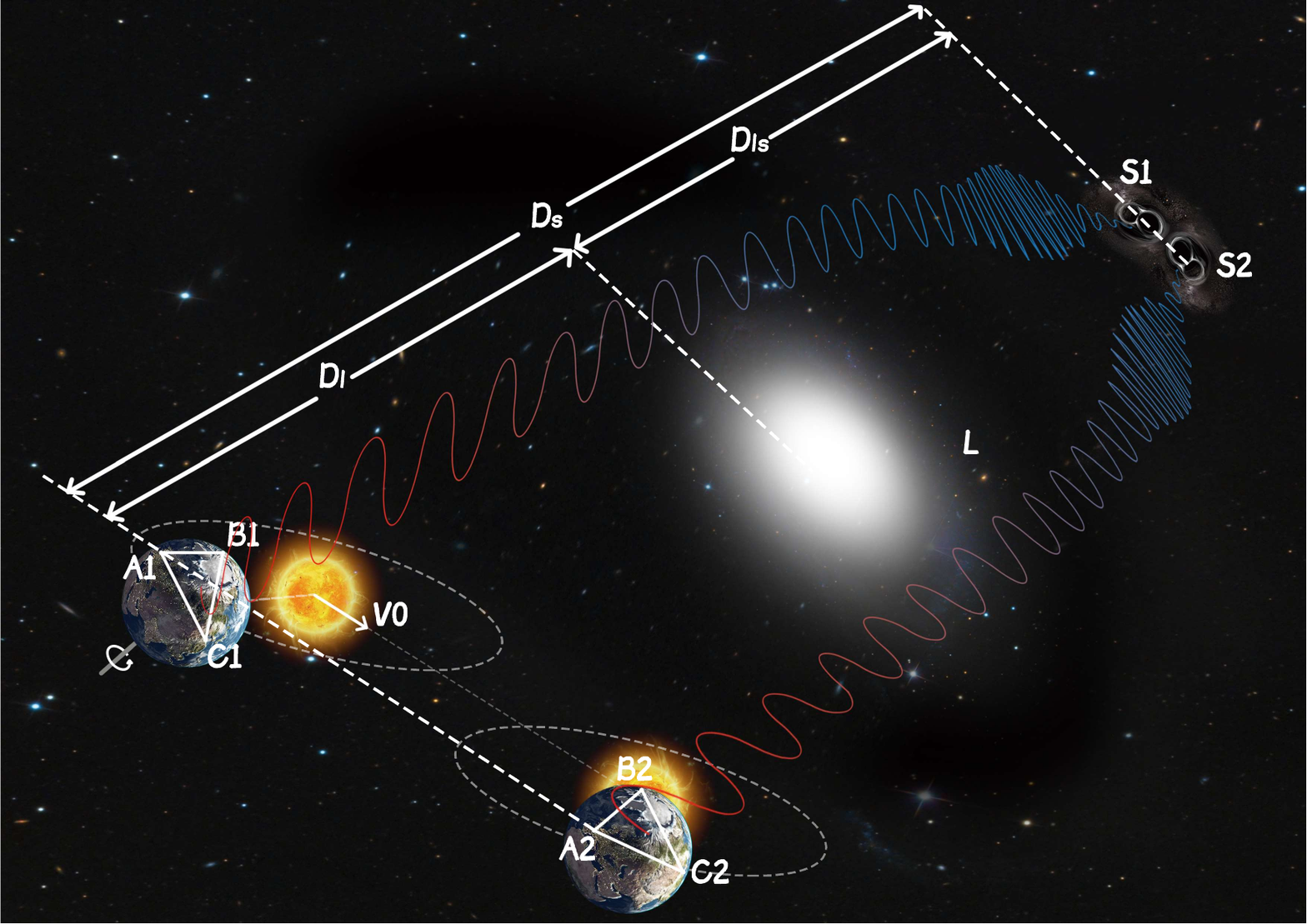}
\caption{
A schematic diagram for the effective network of ground-based GW detectors for a lensed GW event with double images detected at different time. S$_1$ and S$_2$ are the double images of the event, L is the foreground lens galaxy, and $v_0$ denotes the velocity of the Sun in the Galaxy, $(\rm A_1,B_1,C_1$) and $(\rm A_2,B_2,C_2)$ represent the locations of three GW detectors in the space when they receive the signal from S$_1$ and S$_2$, respectively. Due to the rotation of the Earth, usually the baselines $\rm A_2B_2$, $\rm B_2C_2$, and $\rm C_2A_2$ are not parallel with $\rm A_1B_1$, $\rm B_1C_1$, and $\rm C_1A_1$. However, the baselines $\rm A_1A_2$, $\rm B_1B_2$, and $\rm C_1C_2$ are close to, though not exactly, parallel to each other due to the lengths of these baselines are much longer than those of $\rm A_1B_1$, $\rm A_1C_1$, and $\rm B_1C_1$.
}
\label{fig:schematic}
\end{figure}

A ground-based GW detector $i$ ($=$A,B,C,$\cdots$) is at different locations in the space when it receives the signals of different images (total number of $M$) of a lensed GW event because of the rotation and the secular motion of the Earth in the space, which is almost equivalent to a network of $M$ detectors with different antenna pattern function observing a single event (see Appendix). If $N$ detectors on the Earth detected the multiple images of the lensed event, then we have an effective network with $M\times N$ detectors to observe the event. Figure~\ref{fig:schematic} illustrates such a case in which three GW detectors observing two images of a lensed event. The involvement of the new $(M-1)\times N$ detectors and the baselines between image pairs may improve the localization precision of the GW sources significantly and enables the identification of their host galaxies. Therefore, the lensed events detected by the $M\times N$ detectors can be viewed as an independent and complementary probe to solve the Hubble tension problem and constrain the nature of dark energy. 

The improvements in the localization and parameter determinations from the lensed GW signals caused by the effective network could be due to the three factors listed below. First, the effective network has more ``detectors'' ($M\times N$) to measure the GW signals and thus lead to the enhancement of the signal-to-noise ratio (S/N). Second, the rotation of the Earth leads to different directions of the detectors with respect to the signals, and thus leads to different pattern functions at the time of detecting different images. This enhances the localization precision and thus also improves the determinations of the intrinsic and extrinsic parameters of the lensed sources. At last, the baseline of two detectors observing different images can be much longer than that of the intrinsic network observing each single image almost at the same time, because of the revolution of the Earth around the Sun and the translation of the Sun in the Galaxy (secular motion) during the period of the time-delay between different images. In principle, the enlarged baseline could lead to significant improvement in the localization and parameter estimation. The degeneracy between the intrinsic time-delay(s) of different images due to lensing and that due to the secular parallax prevents the accurate measurements of the parallax, and thus the effect due to the long baselines of the effective network cannot be fully extracted, unless there are independent ways to break the degeneracy. Nevertheless, localization enhancement can still arise from new short baselines between image pairs detected by different GW detectors on Earth (e.g., $\boldsymbol{l}^{\rm E}_{\rm A_1A_2}-\boldsymbol{l}^{\rm E}_{\rm B_1B_2}$ for detectors A and B observing images 1 and 2), because they are not cancelled out due to the Earth rotation (see Appendix~\ref{secA1} for details).

We adopt the Fisher Information Matrix method \citep{2008PhRvD..77d2001V,2013PhRvD..88h4013R} to estimate the performance of such an effective network of detectors for mock lensed GW events. The mock GW events, including mergers of stellar binary black holes (sBBHs), neutron star-black hole binaries (NSBHs), and binary neutron stars (BNSs), are generated according to the merger rate density and its evolution with redshift simulated by \textbf{StarTrack} \citep{2020A&A...636A.104B}, with a calibration to the constraints on the local merger rate densities obtained by LIGO/Virgo observations \citep{2021arXiv211103606T}. Then we obtain the mock samples of lensed sBBHs, NSBHs, and BNSs that can be detected by different sets of future GW detectors (see Appendix C). The celestial coordinates of an event are denoted by two angles, $\theta_{\rm s}$ and $\phi_{\rm s}$ (see the definition in Appendix B), which are randomly assigned for the event on the sky sphere. The lens model is assumed to be non-singular isothermal ellipsoid (NSIE) (see Appendix C). The observed GW signals can be obtained from the waveforms generated by \textbf{PyCBC} \citep{2019PASP..131b4503B} considering the lensing amplification effect. Finally, we estimate the localization error of the event by the Fisher Information Matrix method as
\begin{equation}
\Delta \Omega_{\rm s} = 2 \pi\lvert\sin \theta_{\rm s}\rvert \sqrt{\left\langle\Delta \theta_{\rm s}^{2}\right\rangle\left\langle\Delta \phi_{\rm s}^{2}\right\rangle-\left\langle\Delta \theta_{\rm s} \Delta \phi_{\rm s}\right\rangle^{2}},
\label{eq:SolidAng}
\end{equation}
where $\left\langle\Delta\theta_{\rm s}^{2}\right\rangle$, $\left\langle\Delta\phi_{\rm s}^{2}\right\rangle$, and $\left\langle\Delta\theta_{\rm s}\Delta\phi_{\rm s}\right\rangle$ are the corresponding variances and covariance. Note that the Fisher information matrix method may overestimate the localization uncertainty for high S/N events \citep[e.g.,][]{PhysRevD.89.042004}, and then may be viewed as a conservative estimation. For more details see Appendix~\ref{sec:B}.

\begin{figure*}
\centering
\includegraphics[width=0.73\columnwidth]{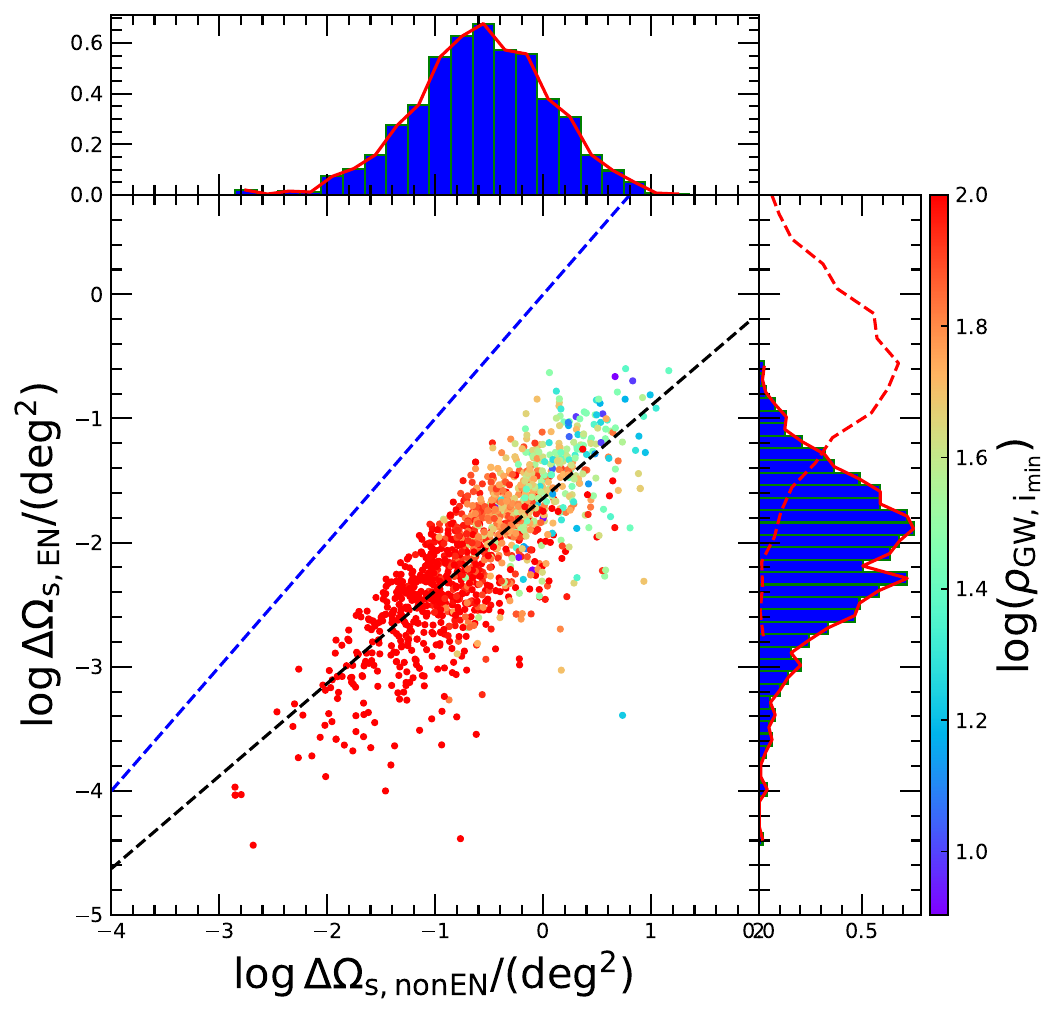}
\includegraphics[width=0.73\columnwidth]{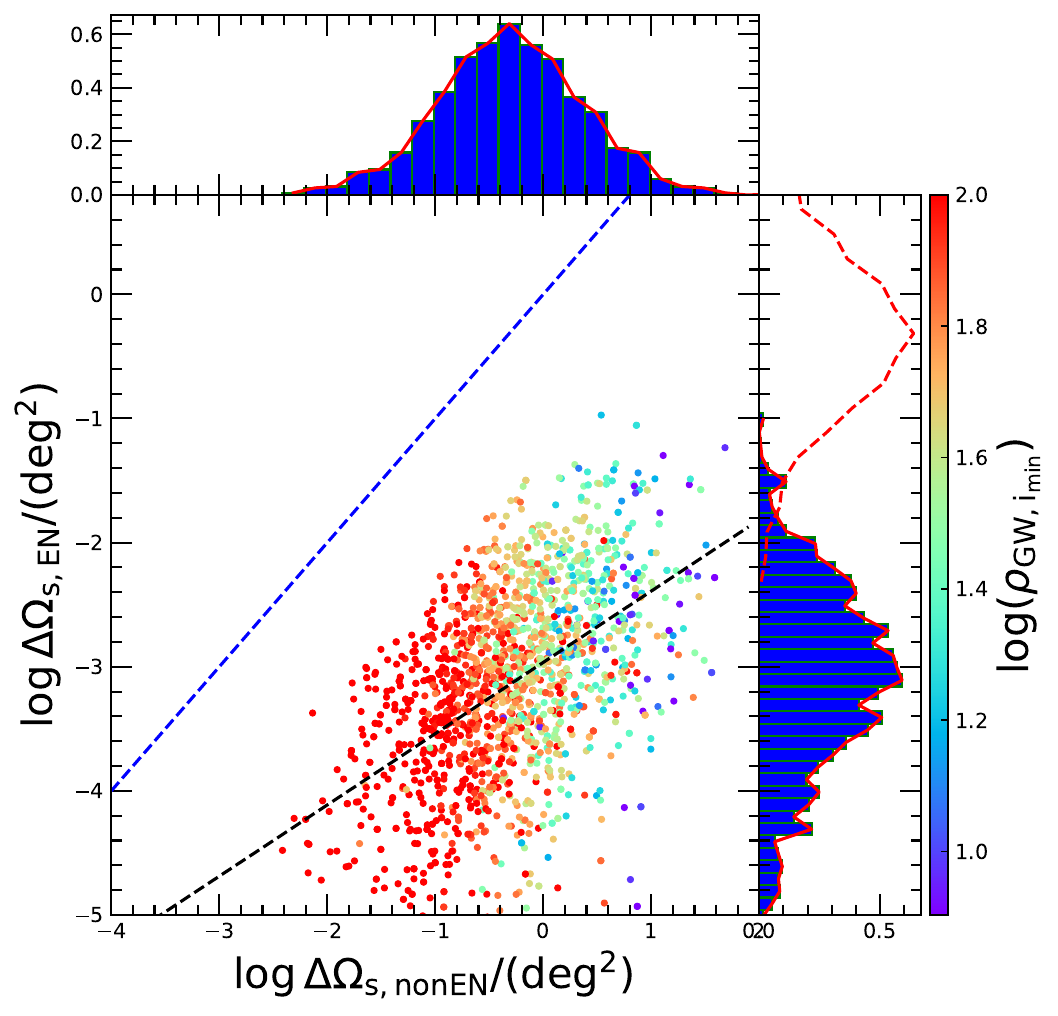}
\caption{
The expected localization uncertainties estimated for mock lensed sBBH GW events considering the effective network (EN) effects due to the motion of the Earth ($\Delta\Omega_{\rm EN}$) against those without considering it ($\Delta\Omega_{\rm nonEN}$). Left and right panels show the cases for those systems with triple and quadruple images, respectively. The color of each symbol represents the S/N of the faintest image with values indicated by the right colour bar. The blue dashed lines represent $\Delta\Omega_{\rm EN}=\Delta\Omega_{\rm nonEN}$. Our sample points are almost all located substantially below the blue dashed lines, which means considering the motion of the Earth in the space can improve the localization precision significantly. The black dotted lines represent the best fits to each sample, i.e., $\log(\Delta\Omega_{\rm s,EN}/\textrm{deg}^2)=0.80\log(\Delta\Omega_{\rm s,nonEN}/\textrm{deg}^2)-1.66$ (left panel) and $\log(\Delta \Omega_{\rm s,EN} /\textrm{deg}^2)=0.57\log(\Delta\Omega_{\rm s,nonEN}/\textrm{deg}^2)-3.07$ (right panel), which suggest the improvement by the effective network to the localization is roughly a factor of $10^{-1.66}$ and $10^{-3.07}$, as also indicated by the top and the right small panels for the one dimension distributions of $\Delta\Omega_{\rm s,EN}$ peaked at $\sim0.01$\,deg$^2$ (or $\sim0.001$\,deg$^2$) and $\Delta\Omega_{\rm s,nonEN}$ peaked at $\sim0.14$\,deg$^2$ (or $\sim0.58$\,deg$^2$). 
}
\label{fig:loc}
\end{figure*}

Figure~\ref{fig:loc} shows the localization errors for the mock lensed GW events (BBHs) with both triple images (left panel) and quadruple images (right panel), detected by three detectors on the Earth with CE \citep{2019BAAS...51g..35R} sensitivity, obtained with consideration of the effective network for lensed sources against those for the same sources without lensing. For the cases with triple images or quadruple images, the distribution of the localization errors obtained by considering the effective network peaks at $\sim0.01$\,deg$^2$ or $\sim0.001$\,deg$^2$ with a scatter of $0.65$\,dex or $0.78$\,dex ($90\%$ confidence level) and has a median of $0.0085$\,deg$^2$ or $0.0008$\,deg$^2$, while that obtained without considering the effect of the effective network peaks at $\sim0.14$\,deg$^2$ or $\sim0.58$\,deg$^2$ with scatter of $0.93$\,dex or $0.86$\,dex and has the median of $0.28$\,deg$^2$ or $0.32$\,deg$^2$. 

For lensed mergers of NSBHs and BNSs, our results are similar as those shown in Figure~\ref{fig:loc}. The improvement of the localization by considering the effective coherent network for lensed sBBH/NSBH/BNS mergers is roughly a factor $\gtrsim10$. Our calculations show that $\sim92.1\%/96.7\%/95.9\%$ of the lensed sources can be localized $10$ times better than the corresponding cases without consideration of the effective network in our calculations. According to these localization error distributions, roughly about $\sim97.4\%/73.2\%/92.8\%$ or $\sim100\%/100\%/100\%$ lensed events with triple images or quadruple images can be localized to areas smaller than $0.1$\,deg$^2$, for lensed sBBH/NSBH/BNS mergers (see Appendix B for more details). Such localization errors for the lensed GW events facilitate the identification of their host galaxies and EM counterparts as discussed below.

\section{Dark Lensed Siren}
\label{sec:dark}

\begin{figure}
\centering
\includegraphics[width=0.49\textwidth]{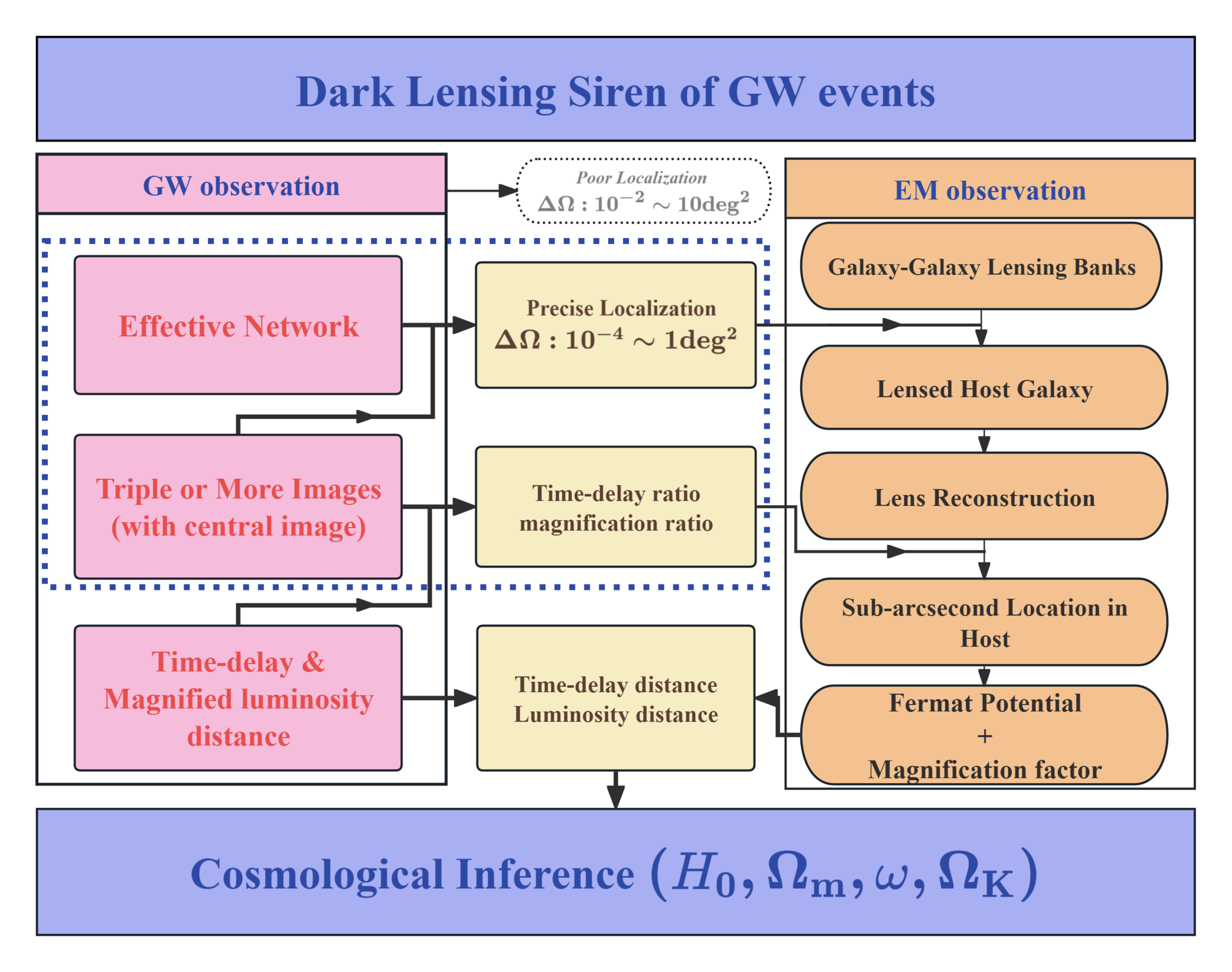}
\caption{
Flowchart of the ``dark lensed siren" method for cosmological inference. In the left and right block, we show the basic recipes needed to obtain or infer from GW and EM (host galaxy) observation, respectively. We use a blue dotted pane to emphasize the key steps in such a method, i.e., (1) precisely localize the GW source via the effective network proposed and find the associated lensed host galaxy; (2) further sub-arcsecond localization of the GW source within the host galaxy via triple or more images (with central image). With the above two key steps, we may constrain the cosmological parameters via the time-delay and luminosity distance measurement from the lensed GW sources, together with the Fermat potential and magnification factor reconstructed by the EM image of the lensed host galaxies.
}
\label{fig:flowchart}
\end{figure}

In this section, we first briefly introduce the main steps of using the ``dark lensed siren" to constrain cosmological parameters by the lensed events with triple/quadruple images. Figure~\ref{fig:flowchart} shows the flowchart of the detailed procedures of this method. First, once the multiple images of the lensed GW signals were detected, we may use the effective network proposed before to localize the GW source within a small sky area. Then the lensed host galaxy may be found in the existed galaxy-galaxy lensing banks or further observations by sky-survey telescopes, such as Chinese Space Survey Telescope (CSST) \citep{2019ApJ...883..203G}, Euclid \citep{2013LRR....16....6A}, and Nancy Grace Roman Space Telescope (RST) \citep{2013arXiv1305.5422S}. Once the associated host galaxy is determined, we may reconstruct the lens mass distribution by the image of the host galaxy (i.e., ring or arc). With the time-delay and magnification ratios accurately measured from the GW signals, we can further localize the GW source in sub-arcsecond precision in the host galaxy and thus determine the Fermat potential difference and magnification factor of the GW source. Finally, together with the time-delays and luminosity distance from the GW signals, we obtain constraints on the cosmological parameters such as ($H_0,\Omega_{\rm m},\omega,\Omega_{\rm K}$). Notably, it is easy to observe that the key point of this ``dark lensed siren" is the identification of lensed host galaxies, which can be achieved by the effective network composed of multiple lensed images of GW sources proposed before. On average no more than $1$ strong lensed system can be observed within a sky area of $\lesssim0.1$-$0.12$\,deg$^2$ by the planned future sky surveys, including Euclid, CSST, and RST \citep[e.g,][]{Collett_2015,2022arXiv221009892C,2023arXiv231206239C}. Therefore, once a lensed galaxy is observed in the localization area ($\leq0.1$\,deg$^2$) of a lensed GW event, it almost grants that this galaxy is the corresponding host galaxy of the event, though a mild mismatch may still be possible due to the existence of other lensed galaxy imposters (see Appendix~\ref{d3} for detailed discussions). 

Hereafter we assess the capability of constraining cosmological parameters with ``dark lensed siren" based on the above procedures. First, we estimate the detection number of GW events by mergers of sBBHs, NSBHs, and BNSs according to the lensing probability. The detection threshold of the S/N for GW signals is set to be $8$ with consideration of the lensing magnification. We assume the LIGO A+ and Voyager \citep{2020CQGra..37p5003A} could be online to work in 2027 and 2030, respectively, and CE \citep{2019BAAS...51g..35R} could be ready for observation in 2035. More details can be seen in Appendix C. The top panel of Figure~\ref{fig:h0} shows the expected number ($N_{\rm len}$) of lensed compact binaries with double images or triple/quadruple images, which can be detected by LIGO A+, LIGO Voyager, and the third generation GW detectors across the period of the year of 2027-2040. As seen from the figure, these detectors are expected to detect $\sim0.167/0.002/0.001$, $2.20/0.12/0.03$, and $13.5/5.98/67.9$ lensed sBBH/NSBH/BNS GW merger events with at least two images per year, respectively. 

Only a fraction of lensed GW events may possess identifiable associated lensed host galaxies. As discussed in \citet{2022arXiv220408732W} and \citet{2022arXiv221009892C}, by future large scale sky-surveys, such as RST/CSST/Euclid may find the lensed host galaxies of $\sim20\%-50\%$ lensed sBBH merger events detected by ground-based GW detectors in their survey areas \citep{2022arXiv220408732W,2022arXiv221009892C}. For BNS and NSBH mergers, we find that this fraction is slightly higher, i.e., $\sim40\%-60\%$, because the lensed BNSs and NSBHs have relatively lower redshifts (see Appendix~\ref{sec:C}). Not all the sky is monitored by these future sky surveys. However, the lensed GW events are interesting and important enough and their localization areas obtained from GW signals are sufficiently small ($\lesssim0.1$\,deg$^2$ for majority of them), so that one may survey each of their localization areas and go deep enough to detect the lensed hosts (as it does not require quick response). Thus, the fraction of the lensed GW events that have detectable lensed hosts is only limited by the recognition of the shape of the lensed host galaxies, for example the Einstein-ring or twisted arcs, which is directly dependent on the angular resolution of the follow-up telescope and/or the relative position of the source to the optical axis of the lens system. We therefore re-estimate the successful host identification fraction of the lensed GW events for mergers of sBBHs, NSBHs, and BNSs and for different GW detectors, respectively (see details in Appendix~\ref{sec:C}), which may be taken as conservative estimations. We find that the detection rate of lensed sBBHs/BNSs/NSBHs by the third generation GW detectors with identifiable host galaxies is $\sim2.18_{-0.35}^{+4.82}/3.02_{-2.27}^{+4.62}/0.46_{-0.24}^{+0.40}$\,yr$^{-1}$ (see Table~\ref{tab:rate} in Appendix).

The middle panel of Figure~\ref{fig:h0} shows the expected number ($N_{\rm len}^{\rm host}$) of those lensed GW events (both mergers of sBBHs, NSBHs, and BNSs) with identifiable host galaxies, considering both the fraction of the lensed mergers that can be localized to within sky areas of $<0.1$\,deg$^2$ by GW signals only, the fraction of those lensed mergers that have identifiable lensed host galaxies estimated above, and the fraction of those events with triple or more images which can be localized with sub-arcsecond precision in the lensed hosts. We emphasize here that only those events with triple or more images and identified host galaxies can be used to constrain the cosmological parameters, because they can be precisely localized in the host galaxy and therefore used to give sufficiently accurate estimate of the corresponding Fermat potential (see Appendix~\ref{d3} for a more detailed discussion). We expect that the total number of such lensed events can be amounted to $\sim30$ at the year of $2040$ by the third generation GW detectors. These events can be used as independent probes to constrain cosmological parameters, especially the Hubble constant accurately. However, the number of such events before 2035 that may be detected by LIGO\,A+ and LIGO Voyager appears negligible as seen from the top panel of Figure~\ref{fig:h0}. 

\begin{figure}[ht]
\centering
\includegraphics[width=0.45\textwidth]{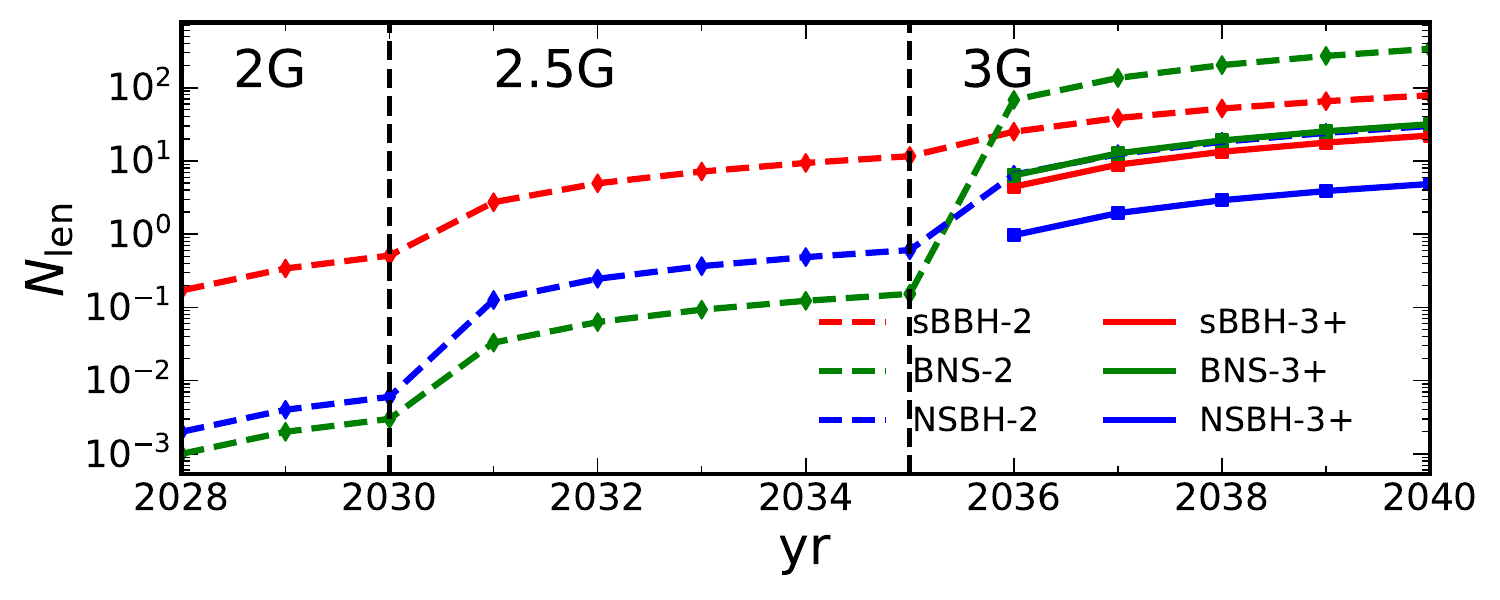}
\includegraphics[width=0.45\textwidth]{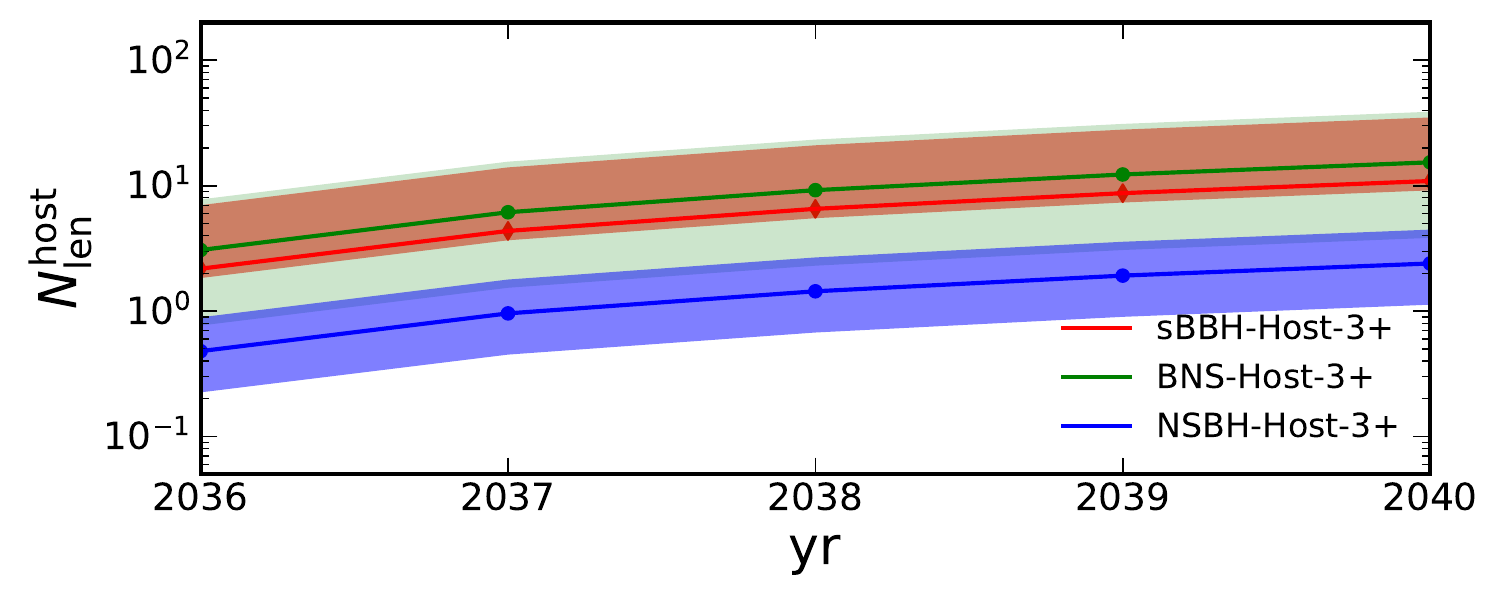}
\includegraphics[width=0.45\textwidth]{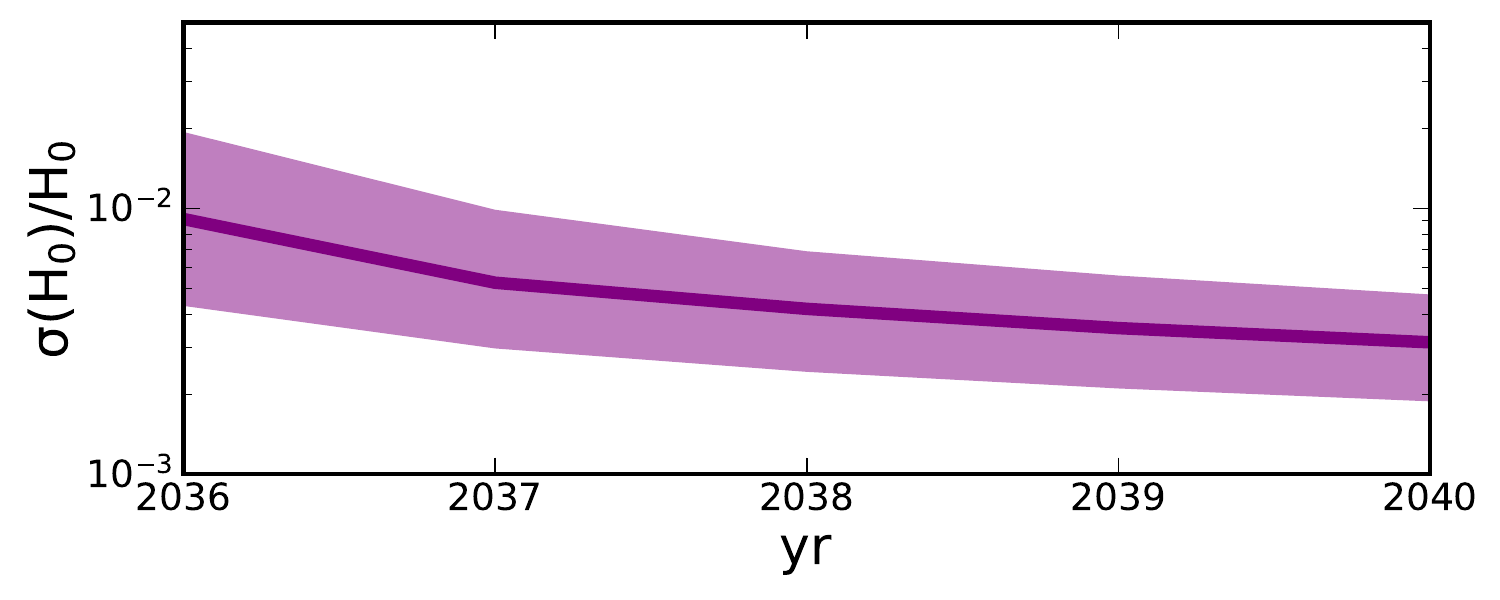}
\caption{
Expected number of lensed GW events ($N_{\rm len}$; top panel), number of the lensed GW event with triple or more images detection with identifiable lensed host galaxies ($N_{\rm len}^{\rm host}$; middle panel), and the corresponding fractional error of the $H_0$ measurement obtained by using these lensed events together (bottom panel) as a function of time. Here we assume that the second/second and a half/third generation (2G/2.5G/3G) GW detectors, i.e., LIGO A+/Voyager/CE, start to work at the year of 2027/2030/2035. In the top panel, the dashed curves with diamond symbols show the results obtained for lensed merger events of sBBHs (red), BNSs (green), and NSBHs (blue) with double images (labelled as sBBH-2, BNS-2, and NSBH-2), respectively, while the solid curves with square symbols show those with triple or more images, labelled as sBBH-3+ (red), BNS-3+ (green), and NSBH-3+(blue), respectively. In the middle and bottom panels, the shaded regions around these curves indicate the uncertainties of the estimates introduced by the uncertainty in the constraints of the local merger rate densities. Note that the numbers for detectable sources shown in the top and middle panel are obtained by averaging those from a large number of realizations.
}
\label{fig:h0}
\end{figure}

Then, we adopt the Markov-Chain-Monte-Carlo (MCMC) process to investigate the performance of the mock lensed GW events with identified host galaxies on constraining the cosmological parameters by using the measurements of their time-delay distance $D_\tau$ and luminosity distance $d_{\rm L}$. The input cosmology model for the simulations of mock events is a flat cold dark matter model ($\Lambda$CDM) with the dimensionless matter density $\Omega_{\rm m}=0.3$, Hubble constant $H_0=70$\,km\,s$^{-1}$\,Mpc$^{-1}$, and cosmological constant $\Omega_\Lambda=0.7$. We assign the Fermat potential uncertainty, line of sight environment uncertainty to $D_{\tau}$ measurements similar as that done in previous studies \citep[e.g.,][]{2017NatCo...8.1148L} (see Appendix~\ref{sec:C} for justification).  For the uncertainties of the intrinsic $d_{\rm L}$ measurements, the weak lensing effect may lead to an error of $\lesssim7.5\%$ to $d_{\rm L}$, where the number of $7.5\%$ is estimated by $\sim0.05z_{\rm s}$ with typical $z_{\rm s}\sim1.5$ \citep{2010CQGra..27u5006S, 2018PhRvD..97f4031Z}, and the reconstruction of host galaxy's magnification may lead to an error of $10-20\%$ \citep{2020MNRAS.498.3395H}. Combining with the errors in the measurements by GW signals, we simply assume that the errors in the intrinsic $d_{\rm L}$ measurements are $\sim20\%$ for demonstration purpose. The bottom panel of Figure~\ref{fig:h0} shows the estimates for the fractional error of $H_0$ constrained by those lensed sBBH/NSBH/BNS events with identified hosts that will be detected at in the future by GW detectors. As clearly seen from this figure, one may obtain the constraint on $H_0$ with a fractional error of $\sim0.90^{+0.90}_{-0.45}\%$ by about one-year observation after 2035. After 5 years of accumulation, the total number of compact binary mergers with identifiable host galaxies will be $\sim20$ and thus $H_0$ can be constrained to an even better precision (e.g., $\sim 0.31_{-0.15}^{+0.16}\%$ for 5 years).

The predicted numbers of lensed GW events with identifiable lensed host galaxies may be affected by the uncertainties in the estimations of their merger rate densities. We note that the preliminary constraints on the local merger rate densities for BNS mergers and NSBH mergers given by LVK O1-O4 observations are $56^{+99}_{-40}$ and $36^{+32}_{-20}$\,yr$^{-1}$\,Gpc$^{-3}$, a factor of $5.7$ and $3.7$ times smaller than those given by LVK O1-O3 observations, respectively, while the local merger rate density for sBBH mergers given by LVK O1-O4 is almost the same as that given by LVK O1-O3 \citep{2025arXiv250708778A}. If we adopt these new constraints, then the detection rate of lensed sBBHs/BNSs/NSBHs by the third generation GW detectors with identifiable host galaxies is $\sim2.18_{-0.23}^{+0.46}/0.54_{-0.38}^{+0.95}/0.13_{-0.07}^{+0.12}$\,yr$^{-1}$ (see Table~\ref{tab:rate} in Appendix). The total number of such events will be $\sim6$ or $\sim15$ after a period of 2-year or 5-year observation, and $H_0$ can be constrained to a precision of $\sim0.90_{-0.48}^{+0.11}\%$ or $\sim0.48_{-0.22}^{+0.09}\%$. In these cases, lensed sBBHs dominate, with a number of $\sim4$ and $\sim11$ after a period of 2-year or 5-year observations, which can solely lead to the Hubble measurement with a precision of $\sim0.94_{-0.13}^{+0.07}\%$ or $\sim0.52_{-0.05}^{+0.03}\%$. Therefore, we conclude that the ``dark lensed siren", even solely the lensed mergers of sBBHs, can put robust constraint on the Hubble constant with a precision of $\lesssim1\%$ after two-year or longer observations by the third generation GW detectors.

We also consider more complex cosmological models, in which the equation state of dark energy ($w$) and curvature ($\Omega_k$) of our universe are set as free constant parameters. Table~\ref{tab:D2} lists the results obtained by the average of $50$ realizations of such mock lensed events. If more than $20$ such lensed events are observed (expected by the year of 2040), $\Omega_{\rm m}$ could be also constrained with a fractional error of $\sim30-35\%$, and $w$ and $\Omega_k$ with an absolute error $\sim0.14-0.16$ and $0.10-0.13$, respectively, assuming the input cosmology model to be $(H_0,\Omega_{\rm m},w,\Omega_k)=(70,0.3,-1,0)$.

We note that the predictions of the constraining power by the lensed compact binary mergers on the cosmological parameters are subject to several uncertainties, but we take our results obtained above on the constraining power of lensed GW events on the Hubble constant and other cosmological parameters as conservative and robust. First, there are some uncertainties in the estimates of the sBBH, NSBH, and BNS merger rate densities and their evolution by population synthesis models, though their local merger rate densities can be calibrated by the constraints already obtained from the LIGO/Virgo observations. Future observations by LIGO/Virgo/KAGRA may detect many more sBBH and BNS merger events at higher redshifts and obtain better constraints on the local merger rate densities and the merger rate density evolution, and thus lead to improved predictions. Second, the EM counterparts of some BNS events may be also directly detected \citep[e.g,][]{2023MNRAS.518.6183M,2017ApJ...848L..12A}, especially in the cases with small localization errors given by GW signals, which thus may lead to a larger fraction (close to $1$) of the lensed events with identifiable hosts if using telescopes with a limiting magnitude of $\sim27-28$ in the H/F bands, such as RST and James Webb Space Telescope (JWST) \citep{2023MNRAS.518.6183M}. Furthermore, for those events with observed EM counterparts, their Fermat potential and magnification may be reconstructed with errors better than that assumed in our analysis. These effects could lead to better constraints than those shown in Figure~\ref{fig:h0} and Table~\ref{tab:D2}.

\section{Conclusions}
\label{sec:con}

In this paper, we first find that the lensed compact binary mergers detected by future GW detectors can be localized to small sky areas ($\sim92.7-100\%$ of them with localizaiton $<0.1$\,deg$^2$) via the effective coherent network composed of GW detectors detecting multiple lensed images of a single event at different locations, which may enable the finding of its host galaxy and/or EM counterparts highly successful. Then, we highlight the importance of the central image of lensed GW signals on the sub-arcsecond localization of GW sources in the host galaxies and find that the detection rate of those lensed sBBHs/BNSs/NSBHs with triple or more images associated with identifiable host galaxies is $\sim2.18_{-0.35}^{+4.82}/3.02_{-2.27}^{+4.62}/0.46_{-0.24}^{+0.40}$\,yr$^{-1}$ or $\sim2.18_{-0.23}^{+0.46}/0.54_{-0.38}^{+0.95}/0.13_{-0.07}^{+0.12}$\,yr$^{-1}$ if adopting the local merger rate densities for sBBHs, BNSs, and NSBHs given by LVK O1-O3 observations or LVK O1-O4 observations. These ``dark lensed siren" with identified host galaxies can provide an independent and unique cosmological probe for cosmological inference. In the case adopting the local merger rate densities given by LVK O1-O4 observations, the lensed sBBHs, probably without EM counterparts, dominate the number of events with identifiable host galaxies, which strengthens the importance of using the ``dark lensed siren'' to infer cosmological parameters. We further find that one may accurately measure the Hubble constant (with precision $\lesssim1\%$) as well as other cosmological parameters by only two-year observations of the third generation GW detectors (ET and CE). 

\section*{acknowledgement}
We thank the anonymous referee for insightful comments. This work is partly supported by the National Natural Science Foundation of China under grant nos.\ 12273050 (YL), 12173001 (QY), 11721303 (QY), 12503001(XG), the National SKA Program of China under grant no.\ 2020SKA0120101 (QY), the National Key Research and Development Program of China under grant nos.\ 2020YFC2201400 (YL), the Strategic Priority Research Program of the Chinese Academy of Sciences under grant no. XDB0550300 (YL), the National Astronomical Observatory of China (grant no. E4TG660101), the Postdoctoral Fellowship Program of CPSF under Grant Number GZB20250735 (ZC). 

\bibliographystyle{aasjournalv7}
\bibliography{ref.bib}

\appendix

\section{Effective Network}
\label{secA1}
 
In this section, we prove the feasibility of the effective network composed of ground based GW detectors observing the multiple images of a lensed GW event at different locations in the space. In such an effective network, note that we take into account the change of pattern function and additional baselines between the images due to the Earth rotation, which has not been mentioned in previous works.

\subsection{Localization of GW sources by a GW detector network}
\label{sec:generic}

The localization of a GW source is extremely poor by a single detector but can be significantly improved by a network of multiple detectors. With the network, the improvement of the source localization and parameter determinations is mainly due to 1) the increased S/N of the signal, 2) the differences in the pattern functions of different detectors \citep{2009LRR....12....2S}, and 3) the accurately measured arrival time differences of the signals detected by different detectors over long baselines, which may enable the triangulation method to better localize the source \citep{2010PhRvD..81h2001W}. The first one may only have a small effect as the S/N increase, about a factor of $\sqrt{N}$, is not large if $N$ is a small number, with $N$ representing the number of detectors in the network with more or less the same sensitivity. The second one may have a moderate effect if the directions of different detectors are significantly different. The third one may have the largest effect if the parallax could be extracted from the time-delay between different images detected at different locations of Earth in the sky.

The measurements of the arrival time differences between GW detectors located at different places are highly precise, normally $\lesssim 1$\,ms. Consider two GW detectors, namely A and B, their spatial separation is denoted by $l_{\rm AB}$. The angle $\theta_{\rm GW}$ between the GW propagation direction $\boldsymbol{\hat{n}}$ and the baseline ${\rm AB}$ can therefore be obtained as
\begin{equation}
\theta_{\rm GW}=\arccos\frac{c\tau_{\rm AB}}{l_{\rm AB}},
\end{equation}
where $c$ is the speed of light, $\tau_{\rm AB}$ ($=\boldsymbol{n}(\theta_{\rm s},\phi_{\rm s})\cdot \boldsymbol{l}_{\rm AB}/c$) is the GW arrival time differences between A and B with $\boldsymbol{n}$ denoting the direction of the source, and $l_{\rm AB}=|\boldsymbol{l}_{\rm AB}|$ with $\boldsymbol{l}_{\rm AB}$ denoting the position vector difference of A and B in the Galactocentric frame. Here the arms of the intereferometer are perpendicular and along the $(x,y)$ axis, $\theta_{\rm s}$ and $\phi_{\rm s}$ are the polar angels defined using the wave propagation direction as the polar axis, $\theta_{\rm s}$ is the angle between the propagation direction and the normal of the detector plane $(x,y)$, and $\phi_{\rm s}$ is the angle in the $(x,y)$ plane measured from the $x$-axis. In principle, the location of a GW source detected by both A and B can be constrained around a circle on the sky defined by the above equation (the circle is denoted by $\odot_{\rm AB}$). If an extra detector C, not located on the extended line of $\rm AB$, also detects the source, then the source may be localized to areas around the intersection points of the two circles in the celestial sphere (i.e., $\odot_{\rm AB}$ and $\odot_{\rm BC}$). Considering a single GW event detected by the network composed of A, B, and C, the GW source can be constrained to two symmetric (small) areas according to the time-delays between the GW signals received by different detectors by taking into account the measurement errors. Given the errors in the measurements of $\tau$, the accuracy of localizing the source increases with increasing the baseline length.

The antenna pattern functions of GW detectors depend on the orientation and localization of GW sources as \citep{2009LRR....12....2S}
\begin{eqnarray}
F_+(t) &=& \sin\zeta\left[a(t)\cos2\psi+b(t)\sin2\psi\right],\\
F_\times(t) &=& \sin\zeta\left[b(t)\cos2\psi-a(t)\sin2\psi\right],
\end{eqnarray}
where $a(t)$ and $b(t)$ are the function of time $t$, $\theta_{\rm s}$, $\phi_{\rm s}$, and the detector location and geometry (including the orientation of the detector's arms with respect to local geographical direction), $\zeta$ is the angle between the interferometer arms ($90^{\rm o}$ for CE and $60^{\rm o}$ for ET), and $\psi$ is the rotation angle of the basis vectors in the transverse plane of the GW propagation direction with respect to the detector's own axes. More details can be seen in \citet{1998PhRvD..58f3001J}. The dependence of the pattern functions on the location of the source is not so sensitive but it can still lead to significant improvement in the localization if applying a network within which different detectors have significantly different detector's planes. One nice example is the improvement to the localization of GW170817 by the addition of the Virgo observations to the LIGO observatories at Hanford and Livingston, though Virgo did not detect the signal with sufficiently high S/N \citep{2017PhRvL.119p1101A}. 
 
\subsection{Localization of lensed GW events by the effective network}

For a strongly lensed GW event, it has multiple images and the ground-based GW detectors may receive multiple images (e.g., $M$ images) at different times and thus at different spatial positions in the space, due to the motion of the Earth in the space within the period of time-delays between different images. In this case, different images of the event are actually coherent and they almost come from the same sky location, with a separation angle $\lesssim 1$\,arcsec on the sky. Therefore, we can take the detectors (A, B, C, ...) that detect $n$ multiple images of a single lensed GW event as an effective network to detect the event. Each GW detector $i$ ($i=$A, B, C, ...), detecting the event at $M$ different times, can be taken as equivalent to $M$ GW detectors $i_1$, $i_2$, ..., $i_M$ detecting the event with long baselines ($\sim$\,AU; see Fig.~\ref{fig:schematic}). The effective network can be considered as the combination of two constellations of detectors roughly at the same time detecting each of the two images, i.e., the generic network A$_1$B$_1$C$_1$... and A$_2$B$_2$C$_2$..., and those detecting different images, e.g., A$_1$A$_2$, B$_1$B$_2$, etc. The detectors detecting the first and second images can have different pattern functions due to the rotation of the Earth, and the directions of the detectors with respect to the source at the time detecting the second images are different from those at the time detecting the first images. 

The two detector constellations have short baselines, which are on the scale of thousands to tens of thousands kilometers and limited by the Earth size, while A$_1$A$_2$, B$_1$B$_2$, etc. can have baselines (typically of $10^7-10^8$\,km for typical time-delays of $1-30$\,days) much longer than those of the generic network (typically of $10^3-10^4$\,km) on the Earth due to the motion of the detectors with the Earth in the space. 

In principle, the long-baselines of A$_1$A$_2$, B$_1$B$_2$, etc. can contribute substantially to the improvement of localization if the time-delay (secular parallax) induced by the long-baseline could be accurately determined. However, the arrival time difference of each detector $i$ in the network that received the signal of the second image compared with that of the first image not only includes the time difference induced by the long baseline, i.e., $\tau_{i_1 i_2}$, but also the lensing time-delay $\tau_{\ell,i_1i_2}$. Here $\tau_{\ell,i_1i_2}$ represents the pure (or intrinsic) time-delay induced by the lensing effect assuming the position of GW detectors is unchanged in the Galactocentric frame. In the real GW detection, $\tau_{\ell,i_1i_2}$ is unknown and it is necessary to solve it out for successfully using the effective network to localize the GW source. Below we show that $\tau_{\ell,i_1i_2}$ can be subtracted from the observed time-delays $T_{i_1i_2}$ if we have three or more GW detectors. 

For simplicity and illustration purpose, we take a strongly lensed GW event with double images as an example. As shown in Figure~\ref{fig:schematic}, the time-delay for a detector $i$ receiving the two images at locations $i_1$ and $i_2$ is
\begin{equation}
T_{i_1i_2}=\tau_{\ell,i_1i_2}+\tau_{i_1i_2},
\label{eq:T_i}
\end{equation}
with $i=$A, B, C, ..., $\tau_{\ell,i_1i_2}$ denoting the intrinsic lensing time-delay for each detector $i$, $\tau_{i_1i_2}$ representing the time-delay induced by the change of spatial geometrical position of detectors. We have 
\begin{equation}
\tau_{i_1i_2}=  \boldsymbol{l}_{i_1i_2}\cdot \boldsymbol{\hat{n}}(\theta_{\rm s},\phi_{\rm s})/c,
\label{eq:tau_i}
\end{equation}
where $\boldsymbol{\hat{n}}= -(\sin\theta_{\rm s}\cos\phi_{\rm s},\sin\theta_{\rm s}\sin\phi_{\rm s},\cos\theta_{\rm s})$ represents the unit vector of the GW propagation direction, and $ \boldsymbol{l}_{i_1i_2}$ represents the position displacement of the detector induced by the motion of the Earth during the period for receiving the two images. This displacement includes those due to the rotation of the Earth itself $\boldsymbol{l}_{i_1i_2}^{\rm E}$, the revolution of the Earth around the Sun $\boldsymbol{l}_{i_1i_2}^{\rm S}$, and the secular motion of the solar system in the Galaxy $\boldsymbol{l}_{i_1i_2}^{\rm G}=\boldsymbol{V}_{0}T_{i_1i_2}$ with $\lvert \boldsymbol{V}_0\rvert=220{\rm km/s}$ \citep{2000gaun.book.....S} , i.e.,  $\boldsymbol{l}_{i_1i_2} = \boldsymbol{l}_{i_1i_2}^{\rm E}+ \boldsymbol{l}_{i_1i_2}^{\rm S}+ \boldsymbol{l}_{i_1i_2}^{\rm G}$. (Note that our Galaxy also has a relative motion with respect to the rest frame of the cosmic microwave background, which can similarly cause a further displacement. Such a displacement could be even larger than that due to the translation of the Sun in the Galaxy, but does not lead to improvement in the localization for the same reason as that discussed for the displacement due to the translation of the Sun below. In principle, this displacement can also be included in our analysis, but for simplicity ignored here.) Note $ \boldsymbol{l}_{i_1i_2}^{\rm E}$ and $ \boldsymbol{l}_{i_1i_2}^{\rm S}$ have periods of $1$\,day and $1$\,year, respectively, and thus $ \boldsymbol{l}_{i_1i_2}^{\rm E}(T_{i_1i_2}+1\,{\rm day})= \boldsymbol{l}_{i_1i_2}^{\rm E}$, $ \boldsymbol{l}_{i_1i_2}^{\rm S}(T_{i_1i_2}+1\,{\rm year})= \boldsymbol{l}_{i_1i_2}^{\rm S}$. The term $\tau_{i_1i_2}$ considers the arrival time difference $\Delta \tau_\parallel$ along the propagation direction of GW signal (even without lensing). The typical distances between GW detectors on the Earth (for example, A and B, or LIGO-Hanford and LIGO-Livingston) is about $l_{\rm AB}\sim10^3-10^4$\,km. Assuming that the angle between the line AB and the GW propagation direction $\hat{n}$ is $\beta$, then $c\Delta \tau_{\parallel}=l_{\rm AB}\cos\beta \lesssim l_{\rm AB}$ and thus $\Delta \tau_\parallel\lesssim3-30$\,ms.

It is necessary to have the information of both $T_{i_1i_2}$ and $\tau_{\ell,i_1i_2}$ for utilizing the long-baseline effect to constrain $\theta_{\rm s}$ and $\phi_{\rm s}$ from Equations~\eqref{eq:T_i} and \eqref{eq:tau_i}. The quantity $T_{i_1i_2}$ can be directly and precisely measured according to the time of the first and second images of the lensed event received by each of the GW detectors with an error $\lesssim 1$\,ms by applying the matched filtering method, due to the high sensitivity and sampling rate of ground-based detectors in the time-domain \citep{2020LRR....23....3A}. The $\tau_{\ell,i_1i_2}$ cannot be directly measured, but one may obtain some information of $\tau_{i_1i_2}$ by cancelling $\tau_{\ell,i_1i_2}$ in the above Equation~\eqref{eq:T_i} as follows. We note that $\tau_{\ell,i_1i_2}$ is different for different detectors (A, B, C, ...) due to their different locations, but the difference can be small for the following reason. The difference between these lensing time-delays $\tau_{\ell,i_1i_2}$ for different detectors includes both the time-delay difference induced by the transverse position difference of detectors (perpendicular to the propagation direction of GW), and the Shapiro time-delay difference due to the difference of GW propagation paths in the lens. For typical lensing systems with $z_\ell\sim 1$ and $z_{\rm s} \sim 1-2$, the GW sources and lens galaxies are located at angular distances of $D_{\rm a}\sim 1-2$\,Gpc, the corresponding Einstein angle $\theta_{\rm E}\sim 1''\sim5\times10^{-6}$\,rad, and the time-delay $\tau (\theta)$ between the two images with an angle separation $\theta$ ($\sim \theta_{\rm E}$) is typically of $\left< \tau \right> \sim 1-30$\,day \citep{2018MNRAS.476.2220L}. The angle change of the source position due to the difference of detector locations is roughly $\Delta\theta\sim\frac{l_{\rm AB}}{D_{\rm a}}$. The intrinsic time-delay difference for GW signals propagating along two paths (having the same $D_\ell$ and $D_{\rm s}$) with $\theta$ ($\sim \theta_{\rm E}$) and $\theta+\Delta\theta$ is then roughly on the order of $\sim \frac{\partial \tau (\theta)}{\partial \theta}  \Delta \theta \sim  \frac{\left <\tau \right>}{\theta_{\rm E}} \Delta \theta \sim  10^{-8}-10^{-10}\,{\rm s}\ll \Delta \tau_\parallel$. Therefore, the differences between $\tau_{\ell,i_1i_2}$ for different detectors are also on this order, which can be neglected.

Consider the case with three ground based GW detectors, e.g., A, B, and C, as that shown in Figure~\ref{fig:schematic}. The effective network can have three baselines, i.e., A$_1$A$_2$, B$_1$B$_2$, and C$_1$C$_2$, and these baselines normally are not parallel to each other because of the rotation of the Earth (normally $T_{\rm A_1A_2}$ is not an integer in unit of a day). Then we can have two equations as shown below by subtracting $\tau_{\ell,i_1i_2}$ in Equation~\eqref{eq:T_i}, i.e.,
\begin{equation}
T_{\rm A_1A_2}-T_{\rm B_1B_2}\simeq \boldsymbol{l}_{\rm A_1A_2}\cdot \boldsymbol{\hat{n}}(\theta_{\rm s},\phi_{\rm s})/c -\boldsymbol{l}_{\rm B_1B_2}\cdot \boldsymbol{\hat{n}}(\theta_{\rm s},\phi_{\rm s})/c, 
\label{eq:T_AB}
\end{equation}
\begin{equation}
T_{\rm A_1A_2}-T_{\rm C_1C_2}\simeq \boldsymbol{l}_{\rm A_1A_2}\cdot \boldsymbol{\hat{n}}(\theta_{\rm s},\phi_{\rm s})/c -\boldsymbol{l}_{\rm C_1C_2} \cdot \boldsymbol{\hat{n}}(\theta_{\rm s},\phi_{\rm s})/c. 
\label{eq:T_AC}
\end{equation}
In the above two Equations, $T_{\rm A_1A_2}$, $T_{\rm B_1B_2}$, and $T_{\rm C_1C_2}$ can be measured accurately from the GW observations, and $\boldsymbol{l}_{\rm A_1A_2}$, $\boldsymbol{l}_{\rm B_1B_2}$, and $\boldsymbol{l}_{\rm C_1C_2}$ can be inferred given the motions of the Earth and the Sun. Therefore, the only two unknown parameters $(\theta_{\rm s}, \phi_{\rm s})$ can be solved from the above two Equations. Equations~\eqref{eq:T_AB} and \eqref{eq:T_AC} can be re-written as
\begin{equation}
\boldsymbol{\mathcal{L}}_{\rm AB} \cdot \boldsymbol{\hat{n}}(\theta_{\rm s},\phi_{\rm s})=1, 
\label{eq:T_ABn}
\end{equation}
\begin{equation}
\boldsymbol{\mathcal{L}}_{\rm AC} \cdot \boldsymbol{\hat{n}}(\theta_{\rm s},\phi_{\rm s})=1, 
\label{eq:T_ACn}
\end{equation}
where $\boldsymbol{\mathcal{L}}_{\rm AB} \equiv \left(\boldsymbol{l}_{\rm A_1A_2} - \boldsymbol{l}_{\rm B_1B_2}\right)/c(T_{\rm A_1A_2}-T_{\rm B_1B_2})$, and $\boldsymbol{\mathcal{L}}_{\rm AC} \equiv \left(\boldsymbol{l}_{\rm A_1A_2} - \boldsymbol{l}_{\rm C_1C_2}\right)/c(T_{\rm A_1A_2}-T_{\rm C_1C_2})$. We can solve these two equations and obtain
\begin{equation}
\boldsymbol{\hat{n}}(\theta_{\rm s},\phi_{\rm s}) = f_1 \boldsymbol{\mathcal{L}}_{\rm AB} +f_2 \boldsymbol{\mathcal{L}}_{\rm AC} + f_3 (\boldsymbol{\mathcal{L}}_{\rm AB} \times \boldsymbol{\mathcal{L}}_{\rm AC}),
\label{eq:secpara}
\end{equation}
with
\begin{equation}
f_1=\frac{\boldsymbol{\mathcal{L}}_{\rm AB}\cdot \boldsymbol{\mathcal{L}_{\rm AC}} -|\boldsymbol{\mathcal{L}}_{\rm AC}|^2 }{(\boldsymbol{\mathcal{L}}_{\rm AB}\cdot \boldsymbol{\mathcal{L}_{\rm AC}})^2- |\boldsymbol{\mathcal{L}}_{\rm AB}|^2|\boldsymbol{\mathcal{L}}_{\rm AC}|^2},\nonumber
\end{equation}
\begin{equation}
f_2=\frac{\boldsymbol{\mathcal{L}}_{\rm AB}\cdot \boldsymbol{\mathcal{L}_{\rm AC}} -|\boldsymbol{\mathcal{L}}_{\rm AB}|^2 }{(\boldsymbol{\mathcal{L}}_{\rm AB}\cdot \boldsymbol{\mathcal{L}_{\rm AC}})^2- |\boldsymbol{\mathcal{L}}_{\rm AB}|^2|\boldsymbol{\mathcal{L}}_{\rm AC}|^2},\nonumber
\end{equation}
and 
\begin{equation}
f_3= \frac{\pm 1}{|\boldsymbol{\mathcal{L}}_{\rm AB}\times \boldsymbol{\mathcal{L}_{\rm AC}}|} \sqrt{1+\frac{|\boldsymbol{\mathcal{L}}_{\rm AB}-\boldsymbol{\mathcal{L}_{\rm AC}}|^2}{(\boldsymbol{\mathcal{L}}_{\rm AB}\cdot \boldsymbol{\mathcal{L}_{\rm AC}})^2- |\boldsymbol{\mathcal{L}}_{\rm AB}|^2|\boldsymbol{\mathcal{L}}_{\rm AC}|^2}}. \nonumber
\end{equation}

According to Equations~\eqref{eq:T_AB} and \eqref{eq:T_AC}, $T_{\rm A_1A_2}-T_{\rm B_1B_2}$ and $T_{\rm A_1A_2}-T_{\rm C_1C_2}$ are roughly on the order of $|\boldsymbol{l}_{\rm A_1A_2}|/c$, $|\boldsymbol{l}_{\rm B_1B_2}|/c$, or $|\boldsymbol{l}_{\rm C_1C_2}|/c$ when $\boldsymbol{l}_{\rm A_1A_2}$, $\boldsymbol{l}_{\rm B_1B_2}$, and $\boldsymbol{l}_{\rm C_1C_2}$ are not parallel to each other. In this case, the long baselines can lead to a significant improvement in the localization according to Equation~\eqref{eq:secpara}. However, for the lensed GW events, $\boldsymbol{l}_{\rm A_1A_2}$, $\boldsymbol{l}_{\rm B_1B_2}$, and $\boldsymbol{l}_{\rm C_1C_2}$ are close to parallel to each other because all the detectors are on the earth and the earth revolution around the Sun dominates over the other sources if the displacements, thus $T_{\rm A_1A_2}-T_{\rm B_1B_2}$ and $T_{\rm A_1A_2}-T_{\rm C_1C_2}$ are approximately on the order of $|\boldsymbol{l}_{\rm A_1A_2}-\boldsymbol{l}_{\rm B_1B_2}|/c\simeq |\boldsymbol{l}^{\rm E}_{\rm A_1A_2}-\boldsymbol{l}^{\rm E}_{\rm B_1B_2}|/c$ and $|\boldsymbol{l}_{\rm A_1A_2}-\boldsymbol{l}_{\rm C_1C_2}|/c\simeq |\boldsymbol{l}^{\rm E}_{\rm A_1A_2}-\boldsymbol{l}^{\rm E}_{\rm C_1C_2}|/c$, much smaller than $T_{\rm A_1A_2} \sim T_{\rm B_1B_2} \sim T_{\rm C_1C_2}$, and also much smaller than $|\boldsymbol{l}_{\rm A_1A_2}|/c$, $|\boldsymbol{l}_{\rm B_1B_2}|/c$, or $|\boldsymbol{l}_{\rm C_1C_2}|/c$. In such a case, the effect due to long baselines (e.g., ${\rm A_1A_2}$, ${\rm B_1B_2}$, or ${\rm C_1C_2}$) is almost cancelled in Equations~\eqref{eq:T_AB} and \eqref{eq:T_AC}, thus does not help significantly for localization as initially expected. The main reason for this cancellation is that the time-delays induced by the change of the spatial geometrical positions of the detectors cannot be directly measured and the indirect measurements are not sufficiently accurate because of their degeneracy with the intrinsic lensing time-delay. Nevertheless, we can  observe that the remaining short-baselines on Earth $|\boldsymbol{l}^{\rm E}_{\rm A_1A_2}-\boldsymbol{l}^{\rm E}_{\rm B_1B_2}|$ and $|\boldsymbol{l}^{\rm E}_{\rm A_1A_2}-\boldsymbol{l}^{\rm E}_{\rm C_1C_2}|$ can still contribute to the localization enhancement of lensed GW sources, although the long-baselines introduced by the Earth revolution around the sun are cancelled. In our following analysis, we also consider the contribution of such new short baselines, e.g., $|\boldsymbol{l}^{\rm E}_{\rm A_1A_2}-\boldsymbol{l}^{\rm E}_{\rm B_1B_2}|$, on the localization precision.

We emphasize here that the improvement of the S/Ns for lensed GW events and the different pattern functions of the detectors in the effective network detecting different images of the events can lead to significant improvement in the localization of the source. To figure out such improvement quantitatively, one may adopt an optimization method, such as the MCMC or mesh-grid finding method, to solve $\theta_{\rm s}$ and $\phi_{\rm s}$. However, many parameters are involved in such a method when considering the lensing of GWs. Below we adopt the simple Fisher matrix method to estimate the localization accuracy of lensed GW events by the effective network for demonstration purposes.

\section{Localization precision}
\label{sec:B}

In this Section, we show details on the estimation of localization precision of lensed GW events by the Fisher information matrix method and Monte Carlo simulations. First, we generate mock samples of lensed GW events with various parameters, including chirp mass $\mathcal{M}_{\rm c}$, mass-ratio $q$, source redshift $z_{\rm s}$, lens redshift $z_{\ell}$, velocity dispersion $\sigma_{v}$, time-delay $\tau_\ell$, and magnification $\mu_1$ and $\mu_2$ by Monte Carlo simulations, following the same procedures described in \citet{2018MNRAS.476.2220L} and \citet{2022arXiv221009892C}. Their orientation angles, i.e., ($\theta_{\rm s}$, $\phi_{\rm s}$, $\theta_{\rm orb}$, $\phi_{\rm orb}$), are uniformly and randomly sampled in the sky, where $\theta_{\rm s}$ and $\phi_{\rm s}$ are the polar angle (ecliptic latitude) and azimuthal angle (ecliptic longitude) of the GW source in ecliptic system, $\theta_{\rm orb}$ and $\phi_{\rm orb}$ represent the polar angle and azimuthal angle of the normal vector of orbital plane of GW sources, respectively. 

For each lensing system with the above parameters fixed, we apply the standard python package \texttt{PyCBC} \citep{2019PASP..131b4503B} with phenomenological model \textbf{IMRPhenomPv3} \citep{2019PhRvD.100b4059K} and \textbf{IMRPhenomPv2-NRTidalv2} \citep{PhysRevD.105.064050} to generate GW waveforms $h_+(f)$, $h_{\times}(f)$ for sBBH and BNS/NSBH in the frequency domain respectively and calculate the lensed GW signals received by each detector. The $m$-th lensed waveform in the frequency domain received by detector $i$ can be expressed as 
\begin{equation} 
h_{i_m}(f)=\sqrt{\mu_1}\left(F^{i_m}_{+}h_{+}(f) + F^{i_m}_{\times} h_{\times}(f)\right)e^{-j{\Phi_{i_m}}}.
\end{equation}
Here $j=\sqrt{-1}$, $F_{a}^{i_{m}}$ are the antenna pattern functions at different positions with $i=$A, B, C, ...,  $m=1$, $2$, $3$... and $a=+$, $\times$.  The exponential factor represents the phase shift due to different location of the GW detectors in this effective network, i.e., 
\begin{equation}
\Phi_{i_m}=2\pi f \boldsymbol{n}(\theta_{\rm s},\phi_{\rm s})\cdot \boldsymbol{r}_{i_m}/c,
\end{equation}
where $\boldsymbol{r}_{i_m}$ is the position vector of the $i$-th GW detector detecting the $m$-th image in the Galactocentric frame. Note that, there should be significant phase shift induced by the time-delay of gravitational lensing, $\tau_{l,i_1 i_2}$. As discussed in Appendix~\ref{secA1}, the long-baseline effect cannot be extracted out due to the degeneracy between the time delays due to parallax and lensing. However, there is still an additional short baseline on Earth for each pair of images, e.g., $\boldsymbol{l}^{\rm E}_{\rm A_1A_2}$ due to the rotation of Earth that can be used for localization. Therefore, we separate the multiple image components in the effective network for the rest detectors ($i$=B, C, ...), but consider the phase shift of each pair of lensed GW signals detected by detector A to take into account the contribution from the baselines between different images, e.g., $\boldsymbol{l}^{\rm E}_{\rm A_1A_2}$. For convenience, we define two whitened signal vectors to represent the two components (similarly, four components in the quadruple image case) in the effective detector network, i.e.,
\begin{equation}  
\hat{\boldsymbol{d}}_{m}(f)=\left(\frac{h_{\rm A_m}(f)}{\sqrt{S_{\rm A}(f)}},\frac{h_{\rm B_m}(f)}{\sqrt{S_{\rm B}(f)}},\frac{h_{\rm C_m}(f)}{\sqrt{S_{\rm C}(f)}}
...\right),
\end{equation}
where $S_i$ denotes the one-sided noise power spectrum of detector $i=$A, B, C, .... The optimal squared S/N is
\begin{equation}
\varrho_{{\rm GW}}^2=\sum_{m}\left(\left.\hat{\boldsymbol{d}}_{m}(f)\right| \hat{\boldsymbol{d}}_{m}(f)\right),
\end{equation}
where the brace bracket denotes an inner product. For any two vector functions $\hat{ \boldsymbol{a}}(f)$ and $\hat{ \boldsymbol{b}}(f)$, we define the inner product as 
\begin{equation}
\left( \hat{ \boldsymbol{a}}(f)\mid \hat{ \boldsymbol{b}}(f) \right)
=2\sum_{n} \int_{f_{\text {min }}}^{f_{\text {max }}}\left\{a_n(f) b_n^{*}(f)+a_n^{*}(f) b_n(f)\right\} d f ,
\end{equation}
where $n$ denotes the $n$-th component of the vector, $f_{\rm min}$ and $f_{\rm max}$ are the lower and upper limits of the frequency of the GW signal detected by the detectors, respectively.
We also define the S/N for the $j$-th image of the lensed event detected by the $i$-th detector as
\begin{equation}
\varrho_{{\rm GW},i_j}^2=\left(\left.\frac{h_{i_j}(f)}{\sqrt{S_i(f)}}\right|\frac{h_{i_j}(f)}{\sqrt{S_{i}(f)}},\right),
\end{equation}
where $h_{i_j}$ is the GW strain of the $j$-th image detected by the $i$-detector. We assume that a lensed GW event is detectable if at least two lensed images can be detected by each GW detector in the network with S/N $\geq 8$. 

The Fisher information matrix for the $m$-th component in the effective network can then be defined as  \citep{1993PhRvD..47.2198F}
\begin{equation}
\Gamma^{m}_{kl}=\left(\left.\frac{\partial{\hat{  \boldsymbol{d}_{m}}(f)}}{\partial p_k} \right| \frac{\partial\hat{ \boldsymbol{d}_{m}}(f)}{\partial p_l}\right),
\end{equation}
where $p_k$ or $p_l$ represents nine system parameters for a binary system with $k,l=1, 2, ..., 9$, i.e., $(m_1,q,t_{\rm c}, \phi_{\rm c}, \theta_\ell, \phi_\ell, \theta_{\rm s}, \phi_{\rm s}, d_{\rm L})$, with $t_{\rm c}$ and $\phi_{\rm c}$ representing the coalescence time and  the GW phase at the coalescence time, $m_{1}$ and $q$ representing the primary mass and the mass ratio, respectively. Then the total Fisher information matrix for the effective network is
\begin{equation}
\Gamma_{kl}=\sum_{m}\Gamma^{m}_{kl}.
\end{equation}
Therefore, the uncertainties for the measurements of these parameters can be given by 
\begin{equation}
\langle\Delta p_k\Delta p_l\rangle=(\Gamma^{-1})_{kl},
\end{equation}
where $(\Gamma^{-1})_{kl}$ is the inverse matrix of $\Gamma_{kl}$. The angular resolution or localization error to the position GW sources can be then estimated by   \citep{2018PhRvD..97f4031Z,2021Resea202114164R}
\begin{equation}
\Delta \Omega_{\rm s}=2 \pi
\lvert\sin \theta_{\rm s}\rvert \sqrt{\left\langle\Delta \theta_{\rm s}^{2}\right\rangle\left\langle\Delta \phi_{\rm s}^{2}\right\rangle-\left\langle\Delta \theta_{\rm s} \Delta \phi_{\rm s}\right\rangle^{2}},
\label{eq:omega}
\end{equation}
where $\left\langle\Delta \theta_{\rm s}^{2}\right\rangle$, $\left\langle\Delta \phi_{\rm s}^{2}\right\rangle$, and $\left\langle\Delta \theta_{\rm s} \Delta \phi_{\rm s}\right\rangle$ are given by $(\Gamma^{-1})_{kl}$. 

\begin{figure}
\centering
\includegraphics[width=0.49\textwidth]{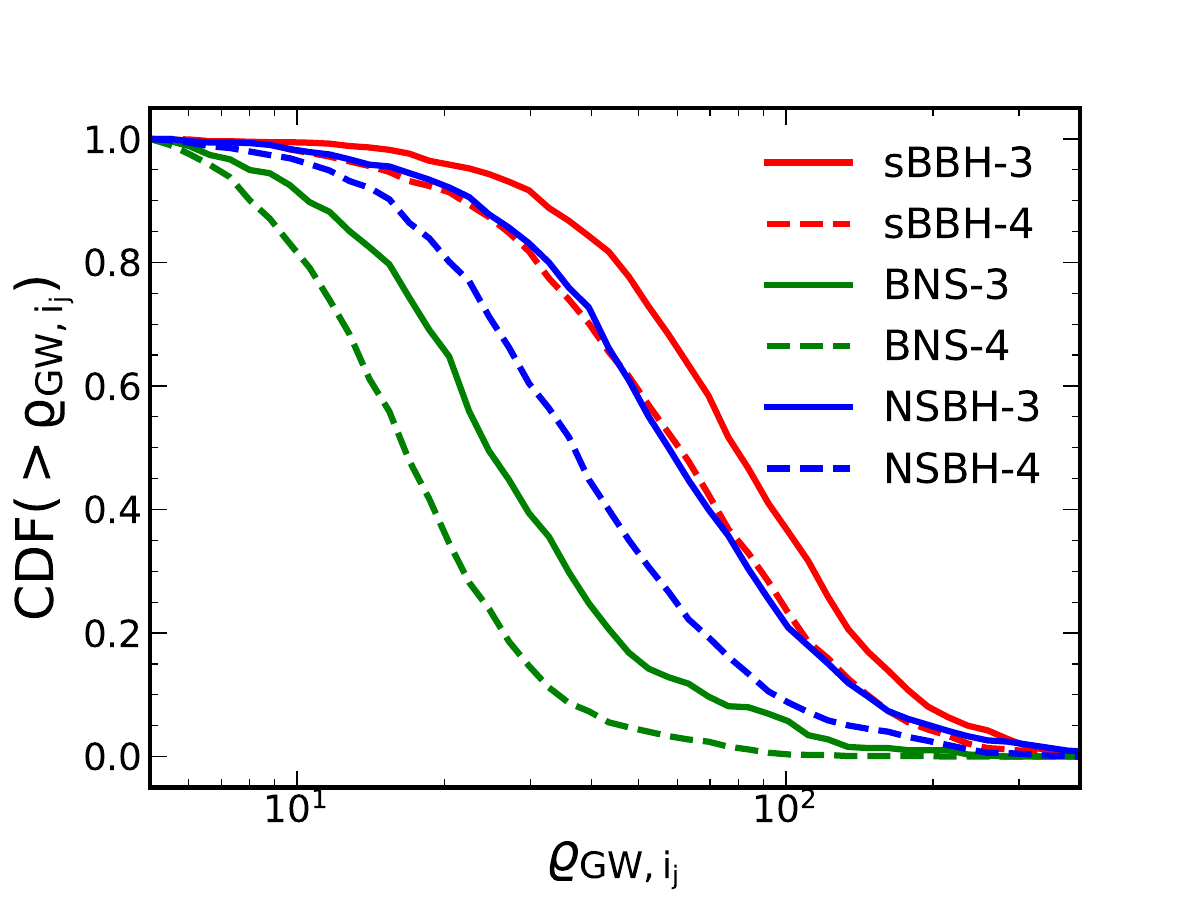}
\caption{
Cumulative probability distribution functions of S/N ($\varrho_{{\rm GW},i_j}$) for all the images of the lensed GW sources ($\rm CDF(>\varrho_{\rm GW})$), normalized to $1$ at $\varrho_{\rm GW}=5$. The red, blue, green solid lines show the results of triple image cases of sBBH, BNS and NSBH, while the dashed lines correspondingly represent the cases of quadruple image. 
}
\label{fig:snr}
\end{figure}

Our calculations show that the mean value of the localization improvement by considering the effective network for lensed sBBH, NSBH, and BNS merger events are roughly a factor of $\sim 29.1$, $30.6$, and $29.7$ (or $\sim 583$, $587$, and $662$) for those cases with triple images (or quadruple) images, respectively. The $10\%$ to $90\%$ ranges for the improvement factor (from low to high) distributions are ($11.2, 96.3$), ($12.2, 205$), and ($10.9, 116$) [or ($84.7, 6.27\times 10^3$), ($90.9, 5.52\times 10^3$), and ($100.1, 5.99\times 10^3$)], respectively (see also the distributions of the localization for lensed compact binary mergers in Figure~\ref{fig:loc} in the main text). Although large scatters do exist in the localization improvement factors of different events due to the differences in their intrinsic/extrinsic parameters, positions, antenna pattern functions, and time-delays, here we conclude that the significant improvement of the localization precision with the effective network mainly results from the following aspects, according to our calculations. First, the involvement of multiple images of lensed GW events (and the additional baseline between two images on Earth) in the effective network enhances the signal-to-noise ratio $\rm S/N$, which contributes to the localization improvement by a factor of $\sim 1.8-5$. Second, change of the pattern function due to the Earth rotation plays a more important role in the improvement of the localization precision, which is about a factor of $\sim 2-15$. Third, due to the magnification bias, the detectable lensed GW events possess relatively larger $\rm S/N$. This will lead to a improvement of the localization by a factor of $\sim 10$. 

The parameter inference based on the Fisher information matrix method may be invalid for low-S/N sources, since the non-Gaussian prior and high-order expansion terms in the signals are neglected in this method \citep{2008PhRvD..77d2001V}; however, as for substantially high-S/N sources, many systems show good agreement in parameter estimation uncertainties between the Fisher information matrix method and the full Bayesian analysis \citep[e.g.,][]{2013PhRvD..88h4013R}. We note that \citet{PhysRevD.89.042004} made a systematic comparison of the $50\%$ credible-interval sky areas determined by the Fisher information matrix method and coherent Bayesian analysis for GW sources with S/N $\geq10$. They found that the Fisher information matrix method tends to overestimate $\Delta\Omega_{\rm s}$ by a median factor of $\sim1.6$, thus offering a rather conservative value. Most of the lensed GW events can be detected with high S/Ns, as shown in Figure~\ref{fig:snr}, and at least about $\sim80\% $ of them have S/N$\geq10$. This suggests that the Fisher information matrix approach used here yields reasonable, albeit slightly conservative, estimates of the localization uncertainties for lensed GW sources, for the purpose of this work, and our main conclusions would not be affected much if adopting the Bayesian approach for each source.

\section{Detection rates}
\label{sec:C}

In this section, we present the details on the estimation of detection rate of lensed GW events with identifiable lensed host galaxies. 

\subsection{Intrinsic merger rate}

The merger rate of the GW events per unit redshift per unit mass of the primary component $m_1$ per unit mass ratio $q$ at redshift from $z_{\rm s}$ to $z_{\rm s}+dz_{\rm s}$ can be described as
\begin{equation}
\frac{d^3\dot{N}_{\rm GW}}{dz_{\rm s}  dm_1dq}=\frac{\mathcal{R}(z_{\rm s},m_1,q)}{1+z_{\rm s}}\frac{dV(z_{\rm s})}{dz_{\rm s}},
\label{source}
\end{equation}
where $\mathcal{R}(z_{\rm s},m_1,q)dm_1 dq$ represents the merger rate density of compact binaries with the primary mass in the range $m_1\rightarrow m_1+ d m_1$ and mass-ratio in the range of $q \rightarrow q+dq$ at redshift $z_{\rm s}$, $dV(z_{\rm s})/dz_{\rm s}$ represents the comoving volume of the universe between redshift $z_{\rm s}$ and $z_{\rm s}+dz_{\rm s}$, and the factor $1/(1+z_{\rm s})$ accounts for the time dilation. For demonstration purpose, we adopt the merger rate density evolution for compact binary object given by \texttt{StarTrack} \citep{2020A&A...636A.104B} to generate mock GW events within $z_{\rm s}\in [0,5]$. For different estimations on the evolution of the merger rate density, the resulting number of the lensed events may only change slightly \citep{2018MNRAS.476.2220L, 2022arXiv221009892C, 2023MNRAS.518.6183M}. Thus the main results in this paper will not be affected much by a different choice of the merger rate density evolution model.

\subsection{Lens model and populations}

Motivated by the existence of a core in the early-type galaxy \citep{1996AJ....111.1889B, 1997ASPC..116..113L}, we consider the Non-Singular Isothermal Ellipsoidal (NSIE) profile as the lensing model. The core size is approximated to be the typical size of the lens systems adopting the simple empirical relation $r_{\rm b}=0.03r_{\rm eff}$ \citep{1997ASPC..116..113L}, where $r_{\rm eff}$ is the effective radius of the lensing galaxy. Here we use the scaling $\langle r_{\rm eff}|\sigma_{v}\rangle$ relation of early-type galaxies obtained from Sloan Digital Sky Survey (SDSS) \citep{2009MNRAS.394.1978H} to estimate $r_{\rm eff}$ from the velocity dispersion $\sigma_{v}$ of the lens galaxy. In galaxy-galaxy strong lensing systems, the central image due to the core-like structure of lens galaxy was rarely considered because of the following two reasons. First, the central image possesses substantially small (de)magnification factor, normally on the scale of $\mu \sim 0.01$. Therefore, the central image is at least $\sim 5$ magnitude fainter than the other images and relatively much more difficult to be detected by EM telescopes. Second, the location of the central image is close to the optical axis of the lens system, which may be strongly contaminated by the emission from the lens galaxy. However, one should note that adding a core-like structure for the lens galaxy is common in the fitting procedure of the member galaxies of a cluster lensing system \citep{2024MNRAS.531.1179X}. Two lensed quasar events with triple images \citep{2004Natur.427..613W} and quintuple images \citep{2005PASJ...57L...7I}, respectively, were indeed detected in the radio band, and the central image is probably due to the core structure.

The central image becomes interesting when considering the lensed GW events. The above two difficulties on the detection of the central image in the cases of EM observations may be naturally avoided for GW detections of lensed GW events. First, the S/N of a lensed GW event is related with the square root of the magnification factor, i.e., $\rm S/N \propto \sqrt{\mu}$. Therefore, the S/N of the central image is only ten times smaller than the other images, which is relatively weak comparing with other images but may be still detectable by future next-generation GW detectors network. Second, the GW signal can get through the lens galaxy without any contamination though it may be affected slightly by the micro-lensing effect from stars or subhalos \citep[e.g.,][]{2025NatAs...9..916S,2025SCPMA..6819512S}. In this sense, we are convinced that the detection of the central image is probable and it can be viewed as promising probes to constrain the central structure of lens galaxies. Therefore, in this paper, we consider the NSIE model and calculate the detection rate of multiple images (i.e., triple/quadruple/quintuple observable images), among which the central image has a relatively substantially smaller magnification factor. This weak image is of great significance in determining the position of the lensed GW event in source plane with sub-arcsecond precision for those with identified lensed host galaxies. 

We model the lens population by adopting the velocity distribution function $dn(\sigma_{v},z_{\ell})/d\sigma_{v}$ given by observations, which is defined so that $dn(\sigma_{v},z_{\ell})$ is the number density of galaxies with $\sigma_{v}$ in the range from $\sigma_{v}$ to $\sigma_{v}+d\sigma_{v}$. The velocity distribution function could be approximated by a simple Schechter function given by \citep{2007ApJ...658..884C, Piorkowska:2013eww},
\begin{equation}
\frac{dn(\sigma_{v},z_{\ell})}{d\ln \sigma_{v}}=n_z \frac{\beta}{\Gamma(\alpha/\beta)}
\left(\frac{\sigma_{v}}{\sigma_{z}}\right)^{\alpha}\exp{\left[-\left(\frac{\sigma_{v}}{\sigma_{z}}\right)^{\beta}\right]},
\label{eq:lens}
\end{equation}
and
\begin{equation}
n_z = n_{0}(1+z)^{\kappa_{n}};\quad \sigma_z = \sigma_{v0}(1+z)^{\kappa_{v}}
\end{equation}
where $\sigma_{v0}$ is the characteristic velocity dispersion, $\alpha$ is the low-velocity power-law index, $\beta$ is the high-velocity exponential cutoff index, ${\Gamma(\alpha/\beta)}$ is the Gamma function, and $(n_0, \sigma_{v0}, \alpha, \beta) = (0.008h^{3} \rm Mpc^{-3}, 161{\rm km\,s^{-1}}, 2.32, 2.67)$. The evolution parameters $\kappa_{n}=-1.18$ and $\kappa_{v}=0.18$ are adopted from the best fit given by \citet{2021MNRAS.503.1319G}.

Once the lens model, lens population, and source population are determined, one can directly estimate the rate for lensed sBBH, BNS, and NSBH mergers (with the primary component mass  $m_1$ and the mass ratio $q$) via \citep[see also][]{2018MNRAS.476.2220L}
\begin{equation}
\begin{aligned}
& \dot{N}_{\rm len} \left(\left.\varrho_{\rm GW}  > \frac{\varrho_{0}}{\sqrt{\mu}} \right| z_{\rm s}\right) =\int_{0}^{\infty} dz_{\rm s}\int_{0}^{\infty} dm_1\int_{0}^{1} dq \int d\boldsymbol{\xi}   \\
&  \tau_{\rm od}(z_{\rm s}) P\left(\left. \varrho_{\rm GW} > \frac{\varrho_0}{\sqrt{\mu}}\right| \boldsymbol{\xi}, z_{\rm s}, m_1,  q\right) \frac{d^3 \dot{N}_{\rm GW} \left(z_{\rm s}, m_1, q\right) }{dz_{\rm s} dm_1 dq},
\end{aligned}
\label{eq:rate}
\end{equation}
with the lensing optical depth 
\begin{equation}
\begin{aligned}
\tau_{\rm od}(z_{\rm s})=&\frac{1}{4 \pi} \int_{0}^{z_{\rm s}} \frac{dV({z_{\ell}})}{dz_{\ell}} dz_\ell \int de P_{\rm e}(e) \\
& \times  \int \frac{dn(\sigma_{v},z_{\ell})}{d\sigma_{v}} d\sigma_{v} S_{\rm cr}(\sigma_{v},z_\ell,z_{\rm s},s,e). 
%
\end{aligned}
\label{eq:tau}
\end{equation}
Here $S_{\rm cr}$ is the cross-section, $P(\varrho_{\rm GW} > \frac{\varrho_{0}}{\sqrt{\mu}}\mid \boldsymbol{\xi},z_{\rm s}, m_1,  q)$ is the conditional probability for a source with parameters of $(\boldsymbol{\xi}=(\theta_{\rm s}, \phi_{\rm s}, \theta_\ell, \phi_\ell),z_{\rm s}, m_1, q)$ that has $\rm S/N$ larger than $\varrho_0$. The value of this probability should be either $1$ or $0$. $P_{\rm e}$ is the probability distributions of the internal NSIE ellipticity. Finally, we can obtain the lensing rates as listed in Table~\ref{tab:rate}. According to these mock samples, we can also generate the distribution of lensing time-delay $\tau_{\ell}$. 

\subsection{Identifiable lensed host galaxies}
\label{d3}

We follow the procedure described in \citet{2022arXiv221009892C} to roughly estimate the fraction ($f^{\rm host}$) of lensed GW events that can have identifiable lensed host galaxies by any survey with any given apparent magnitude limit ($m_{\rm lim}$). Here we argue that the lensed hosts can also be identified by powerful telescopes, such as RST, in addition to the surveys because the accurate localization ($\lesssim 0.06$\,deg$^2$) of the lensed GW events enables direct searches of the lensed host galaxies in the small localization areas. Hence, the detection fraction of lensed hosts is only limited by the recognition of the shape of the lensed host galaxies, for example the Einstein-ring or twisted arcs. In this paper, we chose the optimistic value for $f^{\rm host}$ (e.g., for $m_{\rm lim}\gtrsim 28 $) to estimate the detection rate of lensed GW events with identifiable host galaxies and then assess their performance of constraining the cosmological parameters. In this case, we find that future sky-surveys and/or telescopes, such as Rubin, Euclid, CSST, RST, and JWST \citep{2022arXiv220408732W}, can identify the lensed host galaxies of about $50\%$ of the lensed sBBH/NSBH/BNS mergers detectable by next-generation GW detectors. Table~\ref{tab:rate} also lists the detection rates of those lensed sBBH, NSBH, and BNS mergers that can have identifiable host galaxies.

In the main text, we assume that the lensed galaxy found in the localization area $\Delta\Omega_{\rm s}$ of a lensed GW event is the host of the event if $\Delta \Omega_{\rm s} \leq 0.1$\,deg$^2$. This is reasonable since space sky survey telescopes like Euclid and CSST are predicted to observe $\Pi \sim 10$ lensed galaxies per $1$\,deg$^{2}$ sky area. In practice, we note however that there are still mismatching probabilities that the lensed GW event is not associated with the lensed galaxy found in its localization area, i.e., false alarm probability (FAP), due to the existence of other lensed galaxy impostors. Here we make a rough estimation of this FAP based on Poisson statistics, which can be divided into two cases: (1) A fraction of $f^{\rm host}$ lensed GW events have the unique associated host among all the lensed galaxies observed in the localization area $\Delta\Omega_{\rm s}$. Thus, we may assume that the number of lensed galaxies $k$ in the localization area $\Delta\Omega_{\rm s}$ of an event follows a truncated Poisson distribution $P(k|\lambda_k)$ for $k\geq1$, where $\lambda_{\rm k}= \Pi \Delta\Omega_{\rm s}$ is the expectation of $k$. Then the average FAP for each event can be estimated by ${\rm FAP}\sim \sum_{k=2}^{\infty}P(k|\lambda_{k}=\Pi \Delta\Omega_{{\rm s}})(k-1)/k$. (2) The rest  $1-f^{\rm host }$ lensed GW events do not have identifiable host galaxies, resulting in all the lensed galaxies observed in the sky area of $\Delta\Omega_{\rm s}$ are impostors.  Under this case, the FAP can be estimated by a non-truncated Poisson distribution ${\rm FAP}\sim \sum_{k=1}^{\infty}P(k|\lambda_{\rm k}=\Pi \Delta\Omega_{\rm s})$. By Monte Carlo simulations considering the above two cases, we find that the median FAP for the mock GW events with localization uncertainty shown in Figure~\ref{fig:loc} is about $\rm FAP\sim 4.5_{-3.9}^{+20}\%$, where the upper and lower values represent the $10\%$ and $90\%$ quantiles. Note that in the above estimation of FAP we have implicitly assumed that the host matching is conducted randomly for cases with more than one lensed galaxies in the localization area, which may have large space for further improvement. First, similar with the traditional dark siren cosmology \citep[e.g.,][]{1986Natur.323..310S, 2021ApJ...909..218A}, some physical properties of the host galaxy (e.g., stellar mass) may be used as a weight to reduce the FAP \citep[e.g.,][]{2025arXiv250918243B}. Second, the cross-matching of the time-delay of the host galaxy and GW event may also help in reducing the FAP as done in \citet{2020MNRAS.497..204Y} and \citet{2020MNRAS.498.3395H}. Therefore, we conclude that the FAP is acceptable for practical host identification purpose. 
We also further examine that if adopting a conservative threshold $\Delta \Omega_{\rm s}<0.05$\,deg$^2$ (or $<0.01$\,deg$^2$) in order to reduce the FAP, there are still about $89.9\%/80.1\%/50.5\%$ ($51.5\%/40.0\%/17.3\%$) or $99.7\%/98.6\%/93.6\%$ ($94.5\%/86.7\%/59.4\%$) of the lensed sBBH/NSBH/BNS events with triple or quadruple images that can have $\Delta \Omega_{\rm s}\leq 0.05$\,deg$^2$ ($0.01$\,deg$^2$), within a factor of $\sim2$ of those with the threshold of $\Delta \Omega_{\rm s}\leq0.1$\,deg$^2$. If only considering those events with $\Delta \Omega_{\rm s}<0.05$\,deg$^2$ or $<0.01$\,deg$^2$, then the FAPs would decrease to $\sim 3.9_{-3.3}^{+14}\%$ and $\sim 1.6_{-1.3}^{+5.0}\%$, respectively. In summary, we expect that the results obtained below and in the main text do not qualitatively change much considering the mismatch probability between lensed GW events and host galaxies. 

We emphasize here the importance of having the detection of the central image of the lensed GW signal for its cosmological inference, though the lensing nature of the signal can be revealed by the detection of only two images. Using the lensed GW events for the cosmological inference, it is necessary to determine the Fermat potential with high precision (see Fig.~\ref{fig:flowchart}), which requires sub-arcsecond localization of the lensed events in the source plane and well-reconstruction of the potential map for the host galaxy. In the ``bright lensed siren" method \citep{2017NatCo...8.1148L}, the sub-arcsecond localization in host galaxies may be achieved by direct detecting the lensed EM counterparts (the lensed kilonova and/or afterglow signals). However, the joint rate for such sources is small [substantially less than $0.54+0.13 = 0.67$\,yr$^{-1}$ (see the rate for mergers of BNSs and NSBHs in Section~\ref{sec:dark}) when considering that the preliminary constraints on the local merger rate densities of BNSs and NSBHs by LVK O1-O4 are much lower than previous estimates], which makes the ``bright lensed siren" method on constraining the cosmological parameters perhaps ineffective. When it comes to the ``dark lensed siren", the sub-arcsecond localization can be achieved by the cross-matching of the time-delay and magnification ratio if three or more lensed images can be detected. For example in the triple image case, the ratio of the time delay between image 1 and 2 and that between image 1 and 3, i.e., $\tau_{1,2}/\tau_{1,3}$, and the magnification ratio $\mu_1/\mu_2$ and $\mu_2/\mu_3$ can be precisely measured by GW waveforms and be matched with the reconstruction map from the image of lensed host galaxy, which gives sub-arcsecond localization of the source and Fermat potential measurement with percentage level accuracy. 

We also note that the localization of the lensed events in their host galaxies may or may not be sufficiently accurate for those only having double images. Adopting the singular isothermal spheroid (SIS) profile for the lens model, the two-image cases dominate the lensed events. In these cases, with the reconstructed map of the lensed host, the GW source can be located to a circle in the host galaxy by using the magnification ratio $\mu_{1}/\mu_{2}$. On the circle, the Fermat potential difference between the two images $\delta\phi$ is the same and can be determined by the reconstruction of the potential map of the host galaxy. If adopting the singular isothermal ellipsoid (SIE) lens model, there may be both double-image and quadruple-image cases with the former dominating the rate. For rare quadruple-image cases ($<1\%$ or $0.5$\,yr$^{-1}$ by the third generation GW detectors) with identifiable host galaxies, the event position can be accurately constrained (on sub-arcsecond level) and thus they can be used for the cosmological inference \citep{2020MNRAS.498.3395H}. For the double-image cases, the source can be only determined to on a curve with a substantially wide range of Fermat potential differences and thus the direct application of cosmological inference is difficult. According to the constrained curve, however, one may develop a dark standard siren model to constrain the Hubble constant, similarly as those developed for un-lensed sBBH mergers  \citep{1986Natur.323..310S, 2019ApJ...876L...7S, 2021ApJ...909..218A}, which is beyond the scope of this work. In the present work, we find the detection rate of lensed GW events by the third generation GW detectors with three or more images is substantially enhanced (about $\sim 10$\,yr$^{-1}$) (due to the inclusion of the central images) and therefore helps in localizing the GW sources by the effective network and further sub-arcsecond localization of them in their identified lensed host galaxies. Therefore, we conclude that the method of ``dark lensed siren" involving the type III \citep{2021ApJ...908...97L, 2021ApJ...923L...1J} image or more images is much more effective than the ``lensed bright GW siren" and can give independent and accurate cosmological inference beyond the traditional type Ia supernova and CMB measurement. Here type III images represent those sit near the peak of the lens potential, which are usually faint and rare to observe in EM waves due as swamped by the lens galaxy \citep[e.g.,][]{2004Natur.427..613W}.

\begin{figure*}
\centering
\includegraphics[width=1.0\textwidth]{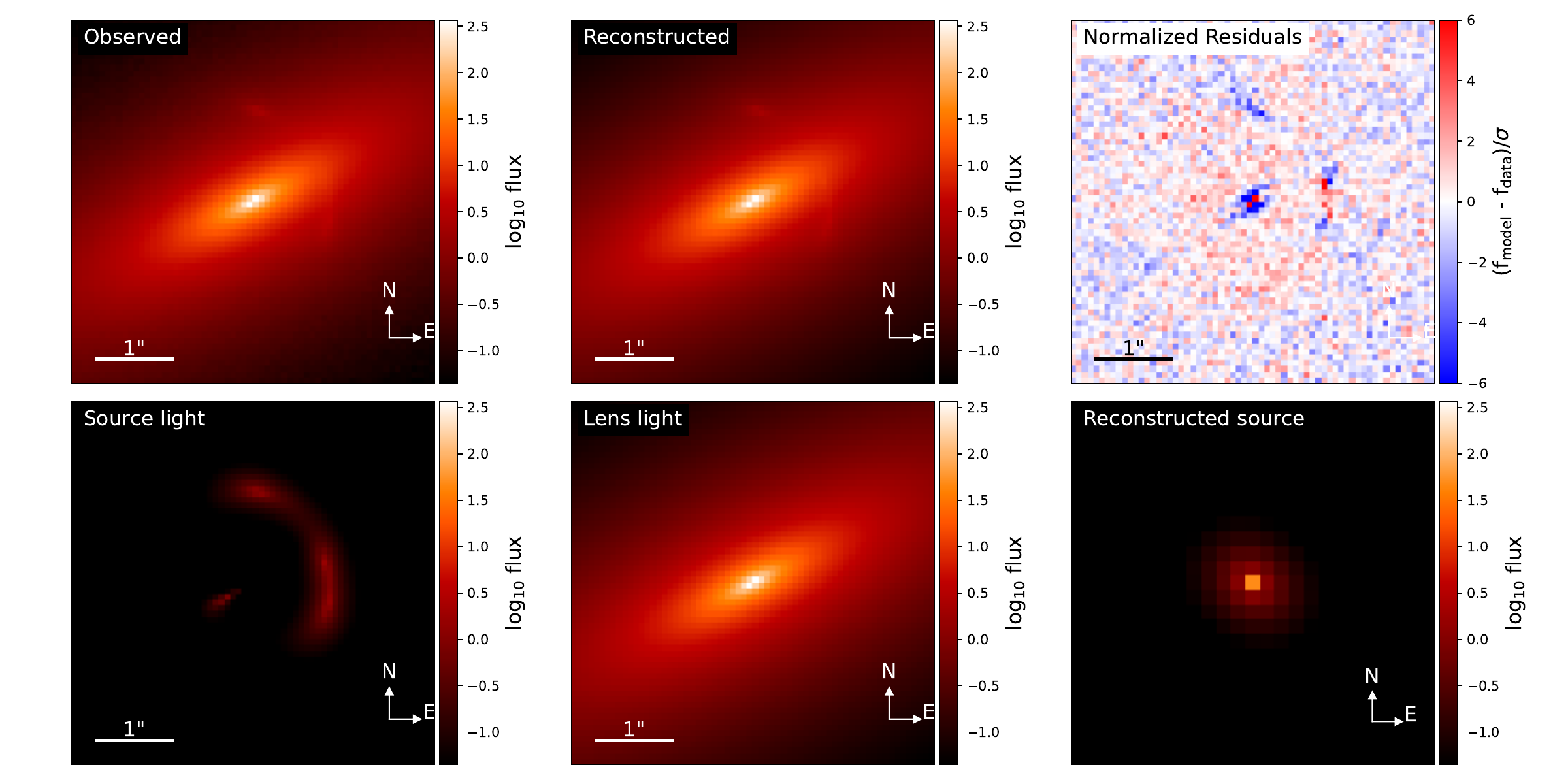}
\caption{Mock image of the lensed system observed by JWST and reconstructed light components of the mock system, with the JWST parameters of readout noise: $13.25\rm e^{-}$, pixel scale: $0.063'$, CCD gain: $1.82 \rm e/ADU$, and zero-point magnitude: $26.47$ mag. The exposure time is set to be $3600\rm s$ per image and the sky brightness is $28.39\rm mag/arcsec^2$. The magnitude of the lensed host galaxy is set to be $M_{\star}\sim 23$ mag, and the reconstructed accuracy on the potential difference map is $\Delta\phi_{i,j}\sim 0.25\%$. Top left panel shows the mock observed system by JWST; top middle panel shows the reconstructed image of the system; bottom left and middle panels show the decomposed observed light from the mock source and lens galaxies, respectively; top right panel shows the residual map after subtracting the reconstructed image from the mock observed system; the bottom right panel shows the reconstructed original image of the source without being lensed. 
}
\label{fig:jwst}
\end{figure*}

To validate the reconstruction power of host galaxies, we also conduct simulations of the images of the lensed host galaxies via the Python package LENSTRONOMY \citep{2015ApJ...813..102B}. We simulate several typical images with parameters of JWST \citep{2023PASP..135d8001R}, including the CCD-gain, readout noise, zero-point magnitude, and sky-background noise. Then we conduct ray-tracing analysis of each image and reconstruct the potential difference map of different images $\Delta\phi_{i,j}$ at each pixel in the source plane. Note that by the method introduced above, the GW source can then be localized within sub-arcsecond precision and therefore find the corresponding pixel in the image of the host galaxy. We find that with even $M_{\star}\sim 25 \rm mag$ host galaxy, $\Delta\phi_{i,j}$ can be constrained with a precision of $\sim 2.5\%$, which is almost comparable with the errors of lensed quasar cases. If the host galaxy is brighter, for example, as shown in Figure~\ref{fig:jwst}, and if the magnitude is $M_{\star}\sim 23$\,mag, we obtain $\Delta\phi_{i,j}\sim 0.25\%$. As typical lensed host galaxies' apparent magnitudes are normally $<25$\,mag, we assign a typical $\Delta\phi_{i,j}$ of $1\%$ for the cosmological inference using Equation~\eqref{eq:cosm}. 

\section{Cosmological parameter inference}
\label{app:cosmology}

\begin{figure}[ht]
\centering
\includegraphics[width=0.45\textwidth]{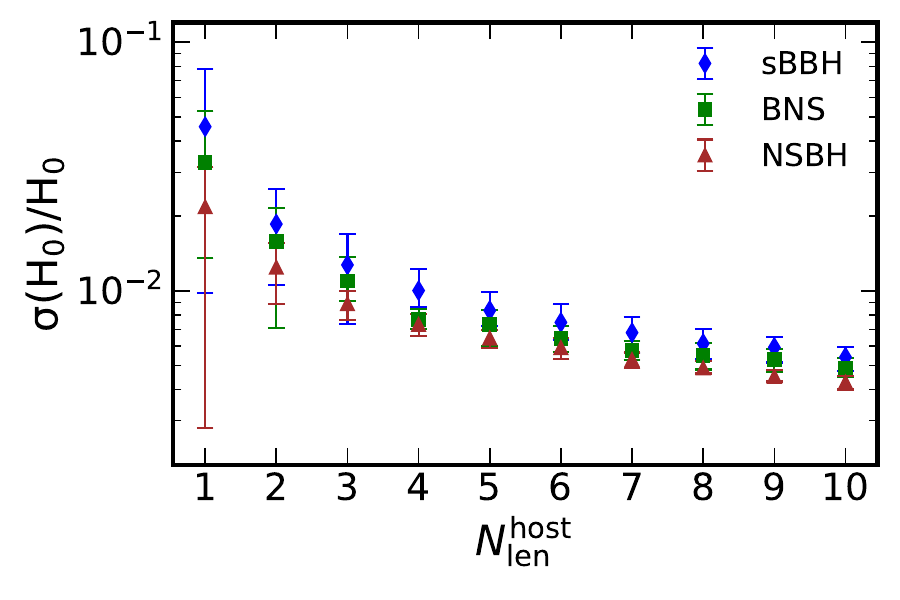}
\includegraphics[width=0.45\textwidth]{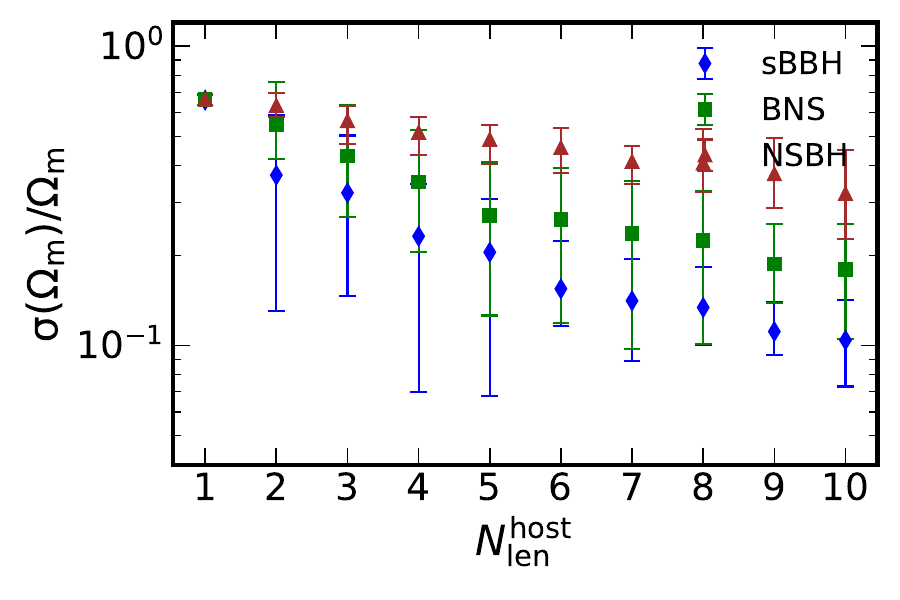}
\caption{
The fractional errors of the constraints on $H_0$ (top panel) and $\Omega_{\rm m}$ (bottom panel) via different numbers of lensed GW sources by assuming the $\Lambda \rm CDM$ model.  The blue, green, and brown symbols show the results obtained from the lensed sBBHs, BNSs, and NSBHs detected by the third-generation (3G) GW detectors (i.e., ET and CE), respectively. The error bars associated with each show the variance (16\%, 84\%) among $50$ realizations.
}
\label{fig:cos2}
\end{figure}

\begin{figure}[ht]
\centering
\includegraphics[width=0.45\textwidth]{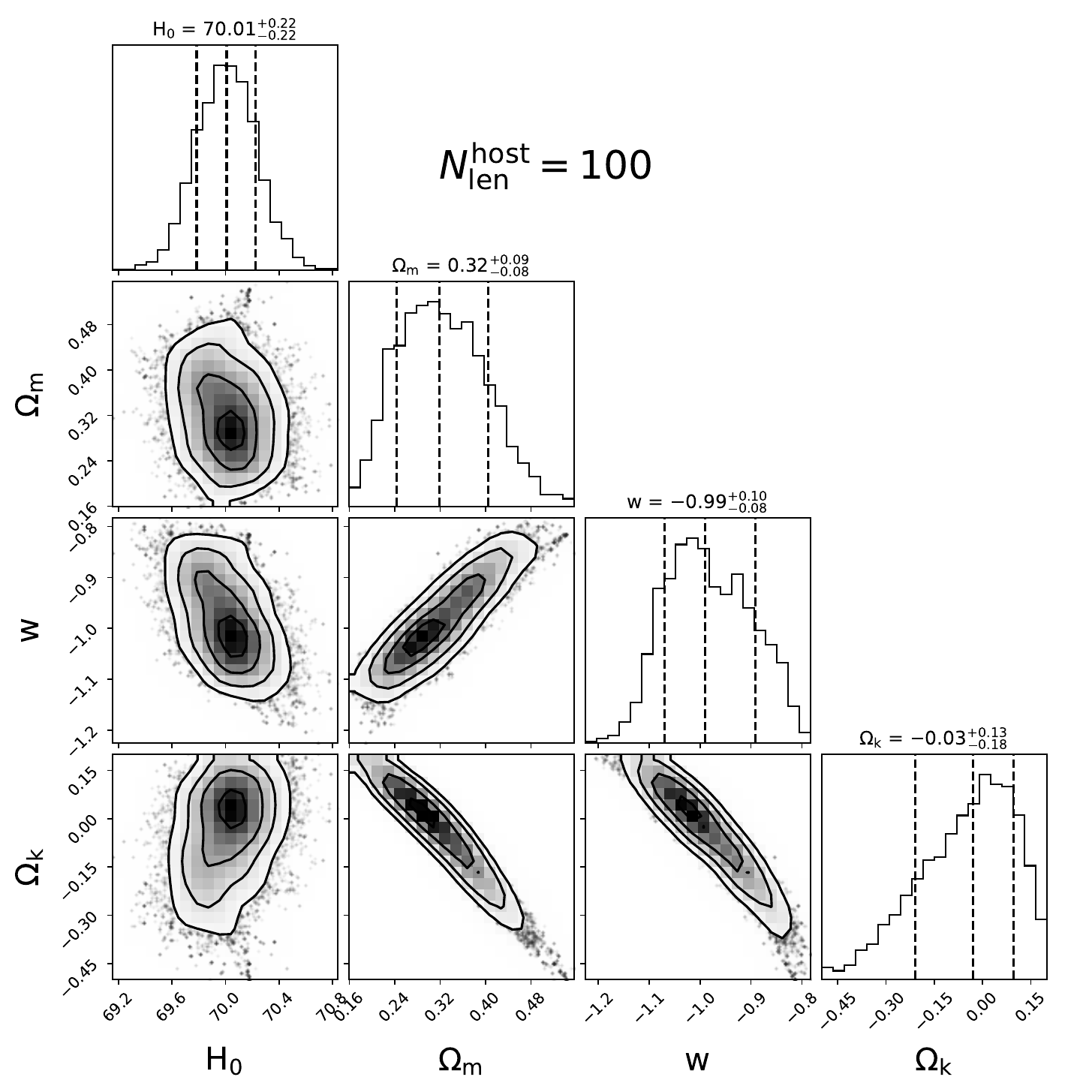}
\caption{
One realization of the cosmological parameter inferences on open $wk$CDM model for $N^{\rm host}_{\rm len}=100$ lensed sBBH mergers with both the GW signals detected by future third generation GW detectors and host galaxies identified by EM telescopes. 
}
\label{fig:cos}
\end{figure}

In this paper, we perform the cosmological parameter inference through the MCMC minimization by using Python package \texttt{emcee} \citep{2013PASP..125..306F} with the $\chi^2$ objective function
\begin{equation}
\chi^2=\sum_{i=1}^{N_{\rm len}^{\rm host}}\frac{(D_{\tau, i}^{\rm th}-D_{\tau,i}^{\rm sim})^2}{\sigma_{D_{\tau, i}}^2}+ \sum_{i=1}^{N_{\rm len}^{\rm host}}\frac{(d_{{\rm L}, i}^{\rm th}-d_{{\rm L},i}^{\rm sim})^2}{\sigma_{d_{{\rm L}, i}}^2},
\label{eq:cosm}
\end{equation}
where $N^{\rm host}_{\rm len}$ is the detection number of the lensed GW events with identifiable host galaxies, $D_{\tau}^{\rm th}$ and $d_{\rm L}^{\rm th}$ are the theoretical time-delay and luminosity distance calculated in the assumed cosmology model, while $D_{\tau}^{\rm sim}$ and $d_{\rm L}^{\rm sim}$ are those inferred from the simulated mock lensed systems with assigned uncertainties, i.e, $\delta d_{\rm L}/d_{\rm L}=20\%$, $\delta \phi/ \phi=1\%$ and $\delta {\rm LOS/LOS}=1\%$, \citep{2017NatCo...8.1148L,2020MNRAS.498.3395H} where  $\delta d_{\rm L}$, $\delta \phi$, and $\delta \rm LOS$ represent the luminosity distance measurement error, the Fermat potential difference, and the line-of-sight (LOS) difference caused by the environment, respectively. Here we note that $\delta \phi/ \phi=1\%$ is based on the assumption of sub-arcsecond localization of GW events and the well-reconstruction of the potential map of the host galaxy as discussed and illustrated above. 

Priors for cosmological parameters, $H_0$, $\Omega_{\rm m}$,  $w$ (for $w$CDM), and $\Omega_k$ (for $wk$CDM), are sampled uniformly in the ranges of $[0,100]$, $ [0,1.2]$, $ [-2,0]$, and $ [-1,1]$, respectively, to obtain the constraints. To avoid the selection bias, we randomly generate $50$ realizations for both the lensed sBBH, NSBH, and BNS mergers, respectively, with which we can obtain the variance of the parameter estimates. 

Figure~\ref{fig:cos2} and Table~\ref{tab:D2} show the average constraining capabilities on the cosmological parameters ($H_0$, $\Omega_{\rm m}$, $w$, $\Omega_k$) under different assumed cosmological models with different numbers of lensed events (sBBHs, NSBHs, or BNSs). As seen from the figure, with a single lensed BNS, NSBH, or sBBH event, one could obtain the constraint on $H_0$ with a fractional error of $\sim1\%-8\%$ by assuming the $\Lambda$CDM model, comparable or even better than that obtained from GW170817, the first multimessenger standard siren detected by LIGO/Virgo \citep{2017Natur.551...85A, 2019NatAs...3..940H, 2020Sci...370.1450D} because of better localization and better measurement on the luminosity distance. But the constraint on $\Omega_{\rm m}$ is not tight and has a large fractional error of $\sim60-70\%$. The constraints can be improved with increasing the number of the lensed events. If the number of the lensed BNS, NSBH, or sBBH events with identified hosts increases to $5$ (or even $20$), then the fractional error on the Hubble constant constraint obtained from them is $\sim0.6\%-1\%$ (or $\lesssim 0.5\%$); the constraint on $\Omega_{\rm m}$ can have a fractional error of $\sim 22\%-42\%$ (or $\sim 6-10\%$) if restricting to the $\Lambda$CDM model. If we relax the cosmology model to the $wk$CDM model with $\Omega_k$ and $w$ being free parameters, and if the total number of lensed BNS, NSBH, or sBBH events with identifiable host galaxies can reach $20$, the constraints on $H_0$ and $\Omega_{\rm m}$ can have a fractional error of $\lesssim0.9\%$ and $\sim 32-40\%$, respectively. Note that in this case the constraints on $w$ and $\Omega_k$ can also reach a considerable precision (see Fig.~\ref{fig:cos} and Tab.~\ref{tab:D2}).

Figure~\ref{fig:cos}, as an example, shows the probability distribution of cosmological parameters, $(H_0,\Omega_{\rm m}, w, \Omega_k)$ inferred from $100$ lensed sBBH mergers detected by future third generation GW detectors (ET and CE) with identified hosts/EM counterparts. Under this circumstance, $H_0$ and $\Omega_{\rm m}$ can be constrained with a fractional error of $\sim 0.31\%$ and $\sim25\%$, respectively, and $w$ and $\Omega_k$ can be constrained with an absolute error of $\sim 0.08$ and $0.13-0.18$, respectively. This again illustrates that the gravitational lensed events by mergers of compact binaries, with host galaxies identified by the lensed hosts, can be taken as prestigious probes to accurately measure the Hubble constant as well as other cosmological parameters.

Compared with the BNSs and sBBHs detected by the third generation GW detectors, those detected by the second generation GW detectors locate at relatively lower redshifts, and they have relatively weaker constraining capability on $\Omega_{\rm m}$, but have a similar capability on constraining $H_0$, given the same number of lensed events. We note here that the constraint on $\Omega_{\rm m}$ is not so tight due to the large uncertainties in the $d_{\rm L}$ measurements ($\sim 20\%$). If one puts more efforts on the lensed host galaxy observation, the accuracy of the reconstruction results  should be improved. Thus, $d_{\rm L}$ can be measured more precisely which may lead to tighter constraints on the matter density ($\Omega_{\rm m}$), the dark energy equation of state ($w$), and even the curvature ($\Omega_k$) of our universe \citep{2019ApJ...873...37L}. 

To close this section, we comment on the current constraints of the Hubble constant $H_0$ from different probes. On the one hand, using the Hubble Space Telescope (HST) observations of 70 long-period Cepheids and the Type Ia supernova, \citet{2022ApJ...938...36R} found that the constraint is $H_0=73.01\pm 0.99$\,km\,s$^{-1}$\,Mpc$^{-1}$. On the other hand, the full-mission Planck measurements of the cosmic microwave background (CMB) anisotropies \citep{2020A&A...641A...5P}, combining information from the temperature and polarization maps and the lensing reconstruction, gives the Hubble constant measurement of $H_0=67.43\pm 0.49$\,km\,s$^{-1}$\,Mpc$^{-1}$, which is $5\sigma$ deviating from the constraints from supernova. This significant deviation is known as the Hubble tension \citep{2024arXiv240612106E,2021CQGra..38o3001D}. As for other probes, the constraints are also diverse. For example, by using the time-delay of the lensed quasars, \citet{2020MNRAS.498.1420W} find $H_0=73.3\pm 1.7$\,km\,s$^{-1}$\,Mpc$^{-1}$, close to the results of supernova measurement. However, observations of Big Bang Nucleosynthesis (BBN) and Baryon Acoustic Oscillation (BAO) \citep{2021PhRvD.103h3533A} suggest that the Hubble constant is a small value, i.e., $H_0= 67.33\pm 0.98$\,km\,s$^{-1}$\,Mpc$^{-1}$, close to the CMB's measurement. By parallel and independent calibration of the Tip of the Red Giant Branch (TRGB) applied to Type Ia supernovae beyond Cepheids, \citet{2019ApJ...882...34F} find $H_0=69.8\pm 0.8$ \,km\,s$^{-1}$\,Mpc$^{-1}$, which lies in between the two groups' measurements mentioned in the Hubble tension problem above. These results indicate the extreme difficulty in resolving the Hubble tension for the potential systematic error of each cosmological probe \citep{2024arXiv240612106E, 2021CQGra..38o3001D}. The landscape of this field may change in the next decade with the accumulation of observations in the above several lines, but the Hubble tension would not be expected to be fully solved in a short period without independent method to measure the Hubble constant, as the measurements of $H_0$ by Cepheids and Type Ia Supernovae have already reached one percent level precision and the measurement by CMB may not be expected to change much as not many all sky CMB surveys are planned in the coming decade. The method proposed in this paper is independent of the above probes, and therefore may help to solve the Hubble tension problem. Furthermore, detecting EM counterparts at high redshifts (e.g., $z\sim 1–2$) is exceptionally challenging, so the conventional bright-lensed GW siren approach demands a longer observational baseline to build up a useful sample. This challenge is further amplified by the significantly lower detection rates of lensed BNSs and NSBHs, compared to sBBHs, when one adopts the local merger rate densities inferred from LVK O1–O4. In addition, as the lensed GW events are located at high redshift, this method may also provide independent and important measurements of the other cosmological parameters. Therefore, it is quite promising for the application of this method in the era of next-generation GW detectors. 

\begin{table*}[ht]
\begin{center}
\begin{minipage}{\textwidth} 
\vskip 0.3cm
\centering
\caption{
Expected rates (yr$^{-1}$) for detectable ($\varrho_{\rm GW}>8$) lensed compact binaries (the upper row for each type of mergers) and the expected rates of those having identifiable lensed host galaxies with small localization error ($\Delta \Omega_{\rm s,EN}\leq0.1$\,deg$^2$) (lower row for each type of mergers) by the third generation GW detectors with three CE design considering the effective network. The lensing rate are calculated from Eq.~\eqref{eq:rate} by using the samples simulated by \texttt{StarTrack}  \citep{2020A&A...636A.104B}, which are scaled by the local merger rate densities obtained from GW observations, i.e., $19_{-3}^{+42} \rm Gpc^{-3} yr^{-1}$ for sBBHs, $320_{-240}^{+490} \rm Gpc^{-3} yr^{-1}$ for BNSs, and $130_{-69}^{+112} \rm Gpc^{-3} yr^{-1}$ for NSBHs, obtained from the LIGO/Virgo observations \citep{2021arXiv211103606T}. Besides, we also list the results in the parenthesis based on the local merger rate density estimation from the preliminary constraints using LVK O4 observations by \citet{2025arXiv250708778A}, i.e., $19^{+4}_{-2}$, $56^{+99}_{-40}$, and $36^{+32}_{-20}$\,yr$^{-1}$\,Gpc$^{-3}$, for sBBHs, BNSs, and NSBHs, respectively.
}
\begin{tabular}{lcccc}
\hline \hline
  & $N_{\rm image}=3$ & $N_{\rm image}=4$  & $N_{\rm image}=5$ & Fraction with central image  \\ \hline
\multirow{2}{*}{sBBH} & $3.60_{-0.57}^{+7.96} ~(3.22_{-0.38}^{+0.76})$ & $0.85_{-0.13}^{+1.88} ~(0.76_{-0.09}^{+0.18})$ & $3\times10^{-3}$ & \multirow{2}{*}{$73\%$}\\
& $1.76_{-0.28}^{+3.89}~(1.57_{-0.19}^{+0.37})$ & $0.42_{-0.07}^{+0.93} ~(0.38_{-0.05}^{+0.09})$ & $\cdots$ \\ \hline
\multirow{2}{*}{BNS} & $1.24_{-0.93}^{+1.90}~(0.22_{-0.16}^{+0.38})$ & $5.11_{-3.83}^{+7.82}~(0.89_{-0.64}^{+1.58})$ & $1\times10^{-4}$ & \multirow{2}{*}{$\sim5.3\%$}\\ 
& $0.46_{-0.35}^{+0.70}~(0.08_{-0.06}^{+0.14})$ & $2.56_{-1.92}^{+3.92}~ (0.45_{-0.32}^{+0.79})$ & $\cdots$ \\ \hline
\multirow{2}{*}{NSBH} & $0.61_{-0.32}^{+0.52}~(0.17_{-0.09}^{+0.15})$ & $0.36_{-0.19}^{+0.31}~(0.10_{-0.56}^{+0.09})$ & $3\times 10^{-4}$ & \multirow{2}{*}{$50\%$}\\
& $0.28_{-0.15}^{+0.24}~(0.08_{-0.04}^{+0.07})$ & $0.18_{-0.09}^{+0.16}~(0.05_{-0.03}^{+0.05})$ & $\cdots$ \\ \hline
%
%
\end{tabular} 
\label{tab:rate}
\end{minipage}
\end{center}
\end{table*}

\begin{table*}
\begin{center} 
\begin{minipage}{\textwidth}
\centering
\caption{The average constraining power  on cosmological parameters by using different numbers of lensed GW sources detected by future third-generation GW detectors. The mock samples for lensed GW events are simulated based on the event rates given by \texttt{StarTrack} \citep{2020A&A...636A.104B}, with calibration to the constraints on the local merger rates of sBBHs and BNSs. Third column lists the total number of the lensed GW events with identifiable host galaxies adopted for obtaining the constraints. Four to seventh columns list the relative errors in the constraints of the Hubble constant and the matter density relative to the critical density, the accuracy of the constraints on $w$ and $\Omega_k$, respectively, obtained by using for $50$ realizations of the lensed events. The input parameters for each model to generate the mock sample are listed below. 
}
\vskip 0.3cm
%
\begin{tabular}{c|cccccc}\hline
%
Events & Cosmology & $N_{\rm len}^{\rm host}$ & $\sigma (H_0)/H_0$ & $\sigma (\Omega_{\rm m})/\Omega_{\rm m}$ & $\sigma (w)$ & $\sigma (\Omega_k)$ \\  \hline   
\multirow{10}{*}{\textbf{sBBH}} & \multirow{4}{*}{$\rm \Lambda CDM$\footnote{$H_0=70$\,km\,s$^{-1}$ and $\Omega_{\rm m}=0.3$.}} & 2 & $3.00\%$ & $47.4\%$ & - & - \\
& & 5& $0.91\%$ & $22.7\%$ & - & -   \\ 
& & 10& $0.56\%$ & $9.67\%$ & - & -   \\ 
& & 100& $0.16\%$ & $2.78\%$ & - & -   \\ \cline{2-7}
& \multirow{3}{*}{$w$CDM \footnote{$H_0=70$\,km\,s$^{-1}$, $\Omega_{\rm m}=0.3$, and $w = -1$.}} &10& $1.13\%$ & $20.7\%$ & 0.12 & -   \\
& & 20& $0.89\%$ & $13.2\%$ & 0.07 & -   \\  
& & 100& $0.30\%$ & $5.10\%$ & 0.03 & -   \\ \cline{2-7}
& \multirow{3}{*}{$wk$CDM\footnote{$H_0=70$\,km\,s$^{-1}$, $\Omega_{\rm m}=0.3$, $w= -1$, and $\Omega_{k}=0.0$.  }}& 20& $0.90\%$ & $32.5\%$ & 0.14 & 0.11   \\
& & 50& $0.50\%$ & $29.6\%$ & 0.12 & 0.10   \\
& & 100& $0.32\%$ & $27.0\%$ & 0.10 & 0.06   \\  \hline
\multirow{10}{*}{\textbf{BNS}} & \multirow{4}{*}{$\Lambda$CDM} & 2 & $1.40\%$ & $57.3\%$ & - & - \\
& & 5& $0.70\%$ & $22.0\%$ & - & -   \\ 
& & 10& $0.50\%$ & $15.6\%$ & - & -   \\ 
& & 100& $0.15\%$ & $4.1\%$ & - & - \\ \cline{2-7}
& \multirow{3}{*}{$w$CDM} &10& $1.10\%$ & $31.9\%$ & 0.18 & -   \\
& & 20& $0.65\%$ & $16.7\%$ & 0.07 & -   \\  
& & 100& $0.27\%$ & $7.70\%$ & 0.03 & -   \\ \cline{2-7}
  & \multirow{3}{*}{$wk$CDM} & 20& $0.67\%$ & $35.1\%$ & 0.14& 0.12  \\
  & & 50& $0.40\%$ & $31.8\%$ & 0.12 & 0.09  \\
  & & 100& $0.28\%$ & $30.3\%$ & 0.11 & 0.09   \\ \hline
\multirow{10}{*}{\textbf{NSBH}} & \multirow{4}{*}{$\Lambda$CDM} & 2 & $1.60\%$ & $59.5\%$ & - & - \\
& & 5& $0.63\%$ & $42.3\%$ & - & -   \\ 
& & 10& $0.41\%$ & $29.4\%$ & - & -   \\ 
& & 100& $0.13\%$ & $7.43\%$ & - & -   \\ \cline{2-7}
&\multirow{3}{*}{$w$CDM} &10& $0.22\%$ & $37.9\%$ & 0.38 & -   \\
  && 20& $0.53\%$ & $26.3\%$ & 0.09 & -   \\   
  && 100& $0.24\%$ & $12.9\%$ & 0.03 & -   \\  
  \cline{2-7}
  & \multirow{3}{*}{$wk$CDM} & 20& $0.60\%$ & $39.4\%$ & 0.16& 0.13  \\
  & &50& $0.35\%$ & $34.0\%$ & 0.12& 0.13  \\
  & & 100& $0.24\%$ & $31.1\%$ & 0.11 & 0.10   \\ \hline  
%
%
\end{tabular}
\label{tab:D2}
\end{minipage}
\end{center}
\end{table*}

\end{document}